# Self-field, radiated energy, and radiated linear momentum of an accelerated point charge

Masud Mansuripur

College of Optical Sciences, The University of Arizona, Tucson, Arizona, USA

**Abstract**. Working within the framework of the classical theory of electrodynamics, we derive an exact mathematical solution to the problem of self-field (or radiation reaction) of an accelerated point-charge traveling in free space. We obtain relativistic expressions for the self-field as well as the rates of radiated energy and linear momentum without the need to renormalize the particle's mass – or to discard undesirable infinities.

**1. Introduction**. The problem of the self-field (or radiation reaction) of an accelerated point-charge has a long and distinguished history [1-14]. Our goal in the present paper is *not* a historical review or a thorough analysis of the many fascinating aspects of this problem. Rather, we aim to present an exact solution of the problem using the classical theory of electrodynamics. We will use Dirac's delta-function to represent the moving point-charge, adhering to an exact mathematical treatment of the $\delta$-function throughout the paper. Within the framework of classical electrodynamics (which is consistent with special relativity), the rates of radiation of energy and linear momentum, given by Eqs.(19) and (20), and the self-field, given by Eq.(27), are exact mathematical results that should be valid for all velocities and accelerations of the point-particle. There will be no need for mass renormalization, nor for sweeping any undesirable infinities under the metaphorical rug. Nevertheless, for reasons that will be explained in Sec.7, the validity of the expression for the self-field given by Eq.(27) remains questionable — at least for particles that move at relativistic velocities.

**2. Retarded potentials**. Within the $(x, y, z)$ Cartesian coordinate system in free space, a point-charge $q$ moves along an arbitrary trajectory $\boldsymbol{r}_p(t) = x_p(t)\hat{\boldsymbol{x}} + y_p(t)\hat{\boldsymbol{y}} + z_p(t)\hat{\boldsymbol{z}}$. The particle's instantaneous velocity and acceleration are $\boldsymbol{v}(t) = \dot{\boldsymbol{r}}_p(t)$ and $\boldsymbol{a}(t) = \ddot{\boldsymbol{r}}_p(t)$, respectively. The charge-density $\rho(\boldsymbol{r}, t)$ and current-density $\boldsymbol{J}(\boldsymbol{r}, t)$ associated with the particle are thus given by

$$\rho(\boldsymbol{r}, t) = q\delta[x - x_p(t)]\delta[y - y_p(t)]\delta[z - z_p(t)] = q\delta[\boldsymbol{r} - \boldsymbol{r}_p(t)]. \tag{1}$$

$$\boldsymbol{J}(\boldsymbol{r}, t) = q\dot{\boldsymbol{r}}_p(t)\delta[\boldsymbol{r} - \boldsymbol{r}_p(t)]. \tag{2}$$

In the Lorenz gauge [15,16], the scalar and vector potentials throughout the spacetime $(\boldsymbol{r}, t)$ are found to be

$$\psi(\boldsymbol{r}, t) = \frac{1}{4\pi\varepsilon_0}\iiint_{-\infty}^{\infty}\frac{\rho(\boldsymbol{r}', t-|\boldsymbol{r}-\boldsymbol{r}'|/c)}{|\boldsymbol{r}-\boldsymbol{r}'|}\mathrm{d}\boldsymbol{r}' = \frac{q}{4\pi\varepsilon_0}\iiint_{-\infty}^{\infty}\frac{\delta[\boldsymbol{r}' - \boldsymbol{r}_p(t')]}{|\boldsymbol{r}-\boldsymbol{r}'|}\mathrm{d}\boldsymbol{r}'. \tag{3}$$

$t' = t - |\boldsymbol{r} - \boldsymbol{r}'|/c$

$$\boldsymbol{A}(\boldsymbol{r}, t) = \frac{\mu_0}{4\pi}\iiint_{-\infty}^{\infty}\frac{\boldsymbol{J}(\boldsymbol{r}', t-|\boldsymbol{r}-\boldsymbol{r}'|/c)}{|\boldsymbol{r}-\boldsymbol{r}'|}\mathrm{d}\boldsymbol{r}' = \frac{\mu_0 q}{4\pi}\iiint_{-\infty}^{\infty}\frac{\dot{\boldsymbol{r}}_p(t')\delta[\boldsymbol{r}' - \boldsymbol{r}_p(t')]}{|\boldsymbol{r}-\boldsymbol{r}'|}\mathrm{d}\boldsymbol{r}'. \tag{4}$$

(See Appendix A for a detailed derivation of the above formulas.) Since the one-dimensional $\delta$-functions that appear implicitly in Eqs.(3) and (4) do *not* have unit areas, they must be scaled in accordance with the standard formula $\delta[f(x')] = \delta(x' - x_0')/|\dot{f}(x_0')|$. Here $x_0'$ is a zero of $f(x')$, while $\dot{f}(x_0')$ is the derivative of $f(x')$ at $x' = x_0'$. From the retardation condition, namely, $t' = t - |\boldsymbol{r} - \boldsymbol{r}'|/c$, we find

$$|\boldsymbol{r} - \boldsymbol{r}'| = c(t - t') \quad\rightarrow\quad \partial_{x'}\sqrt{(\boldsymbol{r} - \boldsymbol{r}')\cdot(\boldsymbol{r} - \boldsymbol{r}')} = -c\partial_{x'}t'$$

$$\rightarrow\quad -\hat{\boldsymbol{x}}\cdot(\boldsymbol{r} - \boldsymbol{r}')/|\boldsymbol{r} - \boldsymbol{r}'| = -c\partial_{x'}t' \quad\rightarrow\quad \partial_{x'}t' = (x - x')/(c|\boldsymbol{r} - \boldsymbol{r}'|) \tag{5}$$

$$\rightarrow\quad \partial_{x'}[x' - x_p(t')] = 1 - \dot{x}_p(t')\partial_{x'}t' = 1 - \frac{\dot{x}_p(t')(x - x')}{c|\boldsymbol{r} - \boldsymbol{r}'|}. \tag{6}$$

Similar expressions are readily obtained for $\partial_{y'}[y' - y_p(t')]$ and $\partial_{z'}[z' - z_p(t')]$ as well. The scalar and vector potentials given by Eqs.(3) and (4) may now be evaluated, as follows:

$$\psi(\boldsymbol{r},t) = \frac{q|\boldsymbol{r}-\boldsymbol{r}_p(t')|^2}{4\pi\varepsilon_0\{|\boldsymbol{r}-\boldsymbol{r}_p(t')|-[x-x_p(t')]\dot{x}_p(t')/c\}\{|\boldsymbol{r}-\boldsymbol{r}_p(t')|-[y-y_p(t')]\dot{y}_p(t')/c\}\{|\boldsymbol{r}-\boldsymbol{r}_p(t')|-[z-z_p(t')]\dot{z}_p(t')/c\}}. \quad (7)$$

$$\boldsymbol{A}(\boldsymbol{r},t) = \frac{\mu_0 q|\boldsymbol{r}-\boldsymbol{r}_p(t')|^2\,\dot{\boldsymbol{r}}_p(t')}{4\pi\{|\boldsymbol{r}-\boldsymbol{r}_p(t')|-[x-x_p(t')]\dot{x}_p(t')/c\}\{|\boldsymbol{r}-\boldsymbol{r}_p(t')|-[y-y_p(t')]\dot{y}_p(t')/c\}\{|\boldsymbol{r}-\boldsymbol{r}_p(t')|-[z-z_p(t')]\dot{z}_p(t')/c\}}. \quad (8)$$

The scalar and vector potentials can also be derived systematically by going back and forth between the spacetime $(\boldsymbol{r},t)$ and the corresponding Fourier domain $(\boldsymbol{k},\omega)$, as discussed in the next section. Surprisingly, it turns out that the Fourier approach yields expressions for $\psi(\boldsymbol{r},t)$ and $\boldsymbol{A}(\boldsymbol{r},t)$ that are *not* identical to those given in Eqs.(7) and (8). We do not understand the reasons behind the disparity, but the potentials obtained using the Fourier method of the next section are elegant and somewhat easier to analyze mathematically. Thus, the discussion in the remainder of the paper will be based on the potentials given in Eqs.(14) and (15).

**3. Computing scalar and vector potentials using Fourier transformation**. In this approach, the charge and current densities of Eqs.(1) and (2) are initially Fourier transformed, as follows:

$$\begin{aligned}\rho(\boldsymbol{k},\omega) &= \int_{-\infty}^{\infty}\rho(\boldsymbol{r},t)\exp[-\mathrm{i}(\boldsymbol{k}\cdot\boldsymbol{r}-\omega t)]\,\mathrm{d}\boldsymbol{r}\mathrm{d}t\\ &= \int_{-\infty}^{\infty} q\delta[\boldsymbol{r}-\boldsymbol{r}_p(t)]\exp[-\mathrm{i}(\boldsymbol{k}\cdot\boldsymbol{r}-\omega t)]\,\mathrm{d}\boldsymbol{r}\mathrm{d}t = q\int_{-\infty}^{\infty}\exp\{\mathrm{i}[\omega t-\boldsymbol{k}\cdot\boldsymbol{r}_p(t)]\}\,\mathrm{d}t. \quad (9)\end{aligned}$$

$$\boldsymbol{J}(\boldsymbol{k},\omega) = \int_{-\infty}^{\infty}\boldsymbol{J}(\boldsymbol{r},t)\exp[-\mathrm{i}(\boldsymbol{k}\cdot\boldsymbol{r}-\omega t)]\,\mathrm{d}\boldsymbol{r}\mathrm{d}t = q\int_{-\infty}^{\infty}\dot{\boldsymbol{r}}_p(t)\exp\{\mathrm{i}[\omega t-\boldsymbol{k}\cdot\boldsymbol{r}_p(t)]\}\,\mathrm{d}t. \quad (10)$$

In the Lorenz gauge, the scalar and vector potentials in the Fourier domain are (see Appendix A):

$$\psi(\boldsymbol{k},\omega) = \varepsilon_0^{-1}\rho_{\text{free}}(\boldsymbol{k},\omega)\left\{\frac{1}{k^2-(\omega/c)^2}+\left(\frac{\mathrm{i}\pi c}{2k}\right)[\delta(\omega-ck)-\delta(\omega+ck)]\right\}. \quad (11)$$

$$\boldsymbol{A}(\boldsymbol{k},\omega) = \mu_0\boldsymbol{J}_{\text{free}}(\boldsymbol{k},\omega)\left\{\frac{1}{k^2-(\omega/c)^2}+\left(\frac{\mathrm{i}\pi c}{2k}\right)[\delta(\omega-ck)-\delta(\omega+ck)]\right\}. \quad (12)$$

Upon inverse Fourier transformation, the scalar potential $\psi(\boldsymbol{r},t)$ is found to be

$$\begin{aligned}\psi(\boldsymbol{r},t) &= (2\pi)^{-4}\int_{-\infty}^{\infty}\psi(\boldsymbol{k},\omega)\exp[\mathrm{i}(\boldsymbol{k}\cdot\boldsymbol{r}-\omega t)]\,\mathrm{d}\boldsymbol{k}\mathrm{d}\omega\\
&= \frac{qc}{(2\pi)^3\varepsilon_0}\int_{t'=-\infty}^{t}\int_{k=0}^{\infty}k^{-1}\sin[ck(t-t')]\int_{\theta=0}^{\pi}2\pi k^2\sin\theta\exp[\mathrm{i}k|\boldsymbol{r}-\boldsymbol{r}_p(t')|\cos\theta]\,\mathrm{d}\theta\mathrm{d}k\,\mathrm{d}t'\\
&= \frac{2qc}{(2\pi)^2\varepsilon_0}\int_{t'=-\infty}^{t}|\boldsymbol{r}-\boldsymbol{r}_p(t')|^{-1}\int_{k=0}^{\infty}\sin[ck(t-t')]\sin[k|\boldsymbol{r}-\boldsymbol{r}_p(t')|]\,\mathrm{d}k\mathrm{d}t'\\
&= \frac{qc}{(2\pi)^2\varepsilon_0}\int_{t'=-\infty}^{t}|\boldsymbol{r}-\boldsymbol{r}_p(t')|^{-1}\\
&\qquad\times\int_{k=0}^{\infty}\cos[ck(t-t')-k|\boldsymbol{r}-\boldsymbol{r}_p(t')|]-\cos[ck(t-t')+k|\boldsymbol{r}-\boldsymbol{r}_p(t')|]\,\mathrm{d}k\,\mathrm{d}t'\\
&= \frac{q}{4\pi\varepsilon_0}\int_{t'=-\infty}^{t}|\boldsymbol{r}-\boldsymbol{r}_p(t')|^{-1}\{\delta[t-t'-|\boldsymbol{r}-\boldsymbol{r}_p(t')|/c]-\underbrace{\delta[t-t'+|\boldsymbol{r}-\boldsymbol{r}_p(t')|/c]}\}\mathrm{d}t'\\
&= \frac{q}{4\pi\varepsilon_0}\int_{t'=-\infty}^{t}|\boldsymbol{r}-\boldsymbol{r}_p(t')|^{-1}\delta[t-t'-|\boldsymbol{r}-\boldsymbol{r}_p(t')|/c]\mathrm{d}t'. \quad (13)\end{aligned}$$

Ignore this $\delta$-function for now, because its argument vanishes only when $t' > t$.

Whereas the first $\delta$-function appearing in the penultimate line of Eq.(13) gives rise to the so-called "retarded" potential(s), the second $\delta$-function, if taken into account, would produce "advanced" potential(s). Generally speaking, the second $\delta$-function in the above equation makes *no* contribution to the electromagnetic (EM) fields, unless one is interested in analyzing the "self-field" of an accelerated point-charge in the particle's immediate vicinity. For the time being, we ignore the contributions of the advanced $\delta$-function, but will resurrect it in Sec.5 when the discussion turns to the topic of radiation reaction.

The retarded $\delta$-function's argument in Eq.(13) vanishes when $t' = t - |\boldsymbol{r} - \boldsymbol{r}_p(t')|/c$, at which point $\partial_{t'}\{t - t' - \sqrt{[\boldsymbol{r} - \boldsymbol{r}_p(t')] \cdot [\boldsymbol{r} - \boldsymbol{r}_p(t')]}/c\} = -1 + \dot{\boldsymbol{r}}_p(t') \cdot [\boldsymbol{r} - \boldsymbol{r}_p(t')]/c|\boldsymbol{r} - \boldsymbol{r}_p(t')|$. We thus have

$$\psi(\boldsymbol{r}, t) = \frac{q}{4\pi\varepsilon_0\{|\boldsymbol{r} - \boldsymbol{r}_p(t')| - [\boldsymbol{r} - \boldsymbol{r}_p(t')] \cdot \dot{\boldsymbol{r}}_p(t')/c\}}. \tag{14}$$

The corresponding expression for the vector potential is similarly found to be

$$\boldsymbol{A}(\boldsymbol{r}, t) = \frac{\mu_0 q\, \dot{\boldsymbol{r}}_p(t')}{4\pi\{|\boldsymbol{r} - \boldsymbol{r}_p(t')| - [\boldsymbol{r} - \boldsymbol{r}_p(t')] \cdot \dot{\boldsymbol{r}}_p(t')/c\}}. \tag{15}$$

The above results manifestly differ from those given by Eqs.(7) and (8), although the two expressions for each potential can be shown to differ perceptibly only at large values of $v/c$, where $\boldsymbol{v} = \dot{\boldsymbol{r}}_p(t')$ is the particle's velocity at time $t'$. In what follows, we shall abandon Eqs.(7) and (8), and rely instead on the scalar and vector potentials given in Eqs.(14) and (15) for purposes of computing the EM field in the vicinity of our moving point-particle.

**4. Electric and magnetic fields and the rate of radiation**. The $E$-field of the accelerated point-charge may now be computed with the aid of the scalar and vector potentials $\psi(\boldsymbol{r}, t)$ and $\boldsymbol{A}(\boldsymbol{r}, t)$. Appendix B provides the details of these calculations. The end result is $\boldsymbol{E}(\boldsymbol{r}, t)$, the electric field at the point $\boldsymbol{r}$ at time $t$, with the location of the point-particle being $\boldsymbol{r}_p(t')$ at the retarded time $t' = t - |\boldsymbol{r} - \boldsymbol{r}_p(t')|/c$. Denoting the separation of $\boldsymbol{r}$ from $\boldsymbol{r}_p(t')$ by the vector $\boldsymbol{R} = \boldsymbol{r} - \boldsymbol{r}_p(t')$, the normalized acceleration and velocity of the particle by $\boldsymbol{\alpha} = \ddot{\boldsymbol{r}}_p(t')/c^2$ and $\boldsymbol{\beta} = \dot{\boldsymbol{r}}_p(t')/c$, and the angle between $\boldsymbol{R}$ and $\boldsymbol{\beta}$ by $\theta$, we will have

$$\boldsymbol{E}(\boldsymbol{r}, t) = -\boldsymbol{\nabla}\psi(\boldsymbol{r}, t) - \partial_t \boldsymbol{A}(\boldsymbol{r}, t) = \frac{q}{4\pi\varepsilon_0}\left[\frac{(1 - \beta^2 + \boldsymbol{R} \cdot \boldsymbol{\alpha})(\hat{\boldsymbol{R}} - \boldsymbol{\beta})}{R^2(1-\beta\cos\theta)^3} - \frac{\boldsymbol{\alpha}}{R(1-\beta\cos\theta)^2}\right]. \tag{16}$$

Similarly, the $B$-field at $(\boldsymbol{r}, t)$ is found to be $\boldsymbol{B}(\boldsymbol{r}, t) = \boldsymbol{\nabla} \times \boldsymbol{A}(\boldsymbol{r}, t) = \hat{\boldsymbol{R}} \times \boldsymbol{E}(\boldsymbol{r}, t)/c$, where $\hat{\boldsymbol{R}} = \boldsymbol{R}/R$ is the unit vector along $\boldsymbol{R}$; see Appendix C. The Poynting vector $\boldsymbol{S}(\boldsymbol{r}, t)$ consists of several terms, with the relevant terms for calculating the radiation rate $\boldsymbol{S}_{\text{rad}}(\boldsymbol{r}, t)$ being those that decline as the square of the distance, $R^2$, from the (retarded) position of the particle. We thus have

$$\boldsymbol{S}(\boldsymbol{r}, t) = \boldsymbol{E} \times \boldsymbol{H} = \boldsymbol{E} \times (\hat{\boldsymbol{R}} \times \boldsymbol{E})/Z_0 = Z_0^{-1}[(\boldsymbol{E} \cdot \boldsymbol{E})\hat{\boldsymbol{R}} - (\boldsymbol{E} \cdot \hat{\boldsymbol{R}})\boldsymbol{E}]. \tag{17}$$

$$\boldsymbol{S}_{\text{rad}}(\boldsymbol{r}, t) \;\rightarrow\; (\boldsymbol{E} \cdot \boldsymbol{E})\hat{\boldsymbol{R}}/Z_0 \;\rightarrow\; \left(\frac{cq^2}{16\pi^2\varepsilon_0}\right)\left[\frac{(1-\beta\cos\theta)^2\alpha^2 + 2(1-\beta\cos\theta)(\boldsymbol{\alpha}\cdot\boldsymbol{\beta})(\boldsymbol{\alpha}\cdot\hat{\boldsymbol{R}}) - (1-\beta^2)(\boldsymbol{\alpha}\cdot\hat{\boldsymbol{R}})^2}{R^2(1-\beta\cos\theta)^6}\right]\hat{\boldsymbol{R}}. \tag{18}$$

In the above equations, $Z_0 = (\mu_0/\varepsilon_0)^{1/2}$ is the impedance of free space. The term $(\boldsymbol{E} \cdot \hat{\boldsymbol{R}})\boldsymbol{E}$ of Eq.(17) makes no contribution to the radiated Poynting vector $\boldsymbol{S}_{\text{rad}}(\boldsymbol{r}, t)$ of Eq.(18). The expression in Eq.(18), when multiplied by $\partial t/\partial t' = 1 - \beta\cos\theta$ [see Eq.(B2) in Appendix B] and integrated over a spherical surface of radius $R$ yields the following rate of EM radiation (see Appendix D for details):

$$\oint_{\text{sphere surface}} (1 - \beta\cos\theta)\boldsymbol{S}_{\text{rad}}(\boldsymbol{r}, t) \cdot \mathrm{d}\boldsymbol{s} = \frac{2}{3}\left(\frac{q^2}{4\pi\varepsilon_0 c^3}\right)\left\{\frac{\ddot{\boldsymbol{r}}_p^2(t')}{(1-\beta^2)^2} + \frac{[\dot{\boldsymbol{r}}_p(t') \cdot \ddot{\boldsymbol{r}}_p(t')]^2}{c^2(1-\beta^2)^3}\right\}. \tag{19}$$

The multiplication of $\boldsymbol{S}_{\text{rad}}(\boldsymbol{r},t)$ by $\partial_{t'}t = 1-\beta\cos\theta$ in the above integrand is necessitated by the fact that the Poynting vector at $(\boldsymbol{r},t)$, being the rate of radiated EM energy per unit (local) time $t$, is obtained as a result of normalizing the energy flux by $\Delta t$, whereas our desired radiation rate should be normalized by $\Delta t'$.

Finally, note that in the non-relativistic regime, where $\boldsymbol{\beta} = \dot{\boldsymbol{r}}_p(t')/c$ is negligible, the radiation rate according to Eq.(19) is simply the product of the constant coefficient $(2q^2/3c^3)/(4\pi\varepsilon_0)$ and the square of the particle's acceleration $\ddot{\boldsymbol{r}}_p(t')$. Other interesting special cases are the case of circular motion, where $\dot{\boldsymbol{r}}_p(t')$ is perpendicular to $\ddot{\boldsymbol{r}}_p(t')$, and the case of oscillatory linear motion, where $\dot{\boldsymbol{r}}_p(t')$ is aligned with $\ddot{\boldsymbol{r}}_p(t')$.

It is also possible to integrate the radiated momentum-density $\boldsymbol{\wp}(\boldsymbol{r},t) = \boldsymbol{S}(\boldsymbol{r},t)/c^2$ over a thin spherical shell of large radius $R$ and infinitesimal thickness $\Delta R = c\Delta t = c(\partial t/\partial t')\Delta t' = c(1-\beta\cos\theta)\Delta t'$. In the limit when $R\to\infty$ and $\Delta t'\to 0$, terms of order $1/R^3$ and $1/R^4$ appearing in Eq.(17) vanish, and the rate of escape of EM momentum per unit time is evaluated from Eq.(18) as follows (see Appendix D for details):

$$\dot{\boldsymbol{p}}(t) = \lim_{\substack{R\to\infty\\ \Delta t'\to 0}} \left\{(\Delta t')^{-1}\int_{\text{sphere surface}} c^{-1}(1-\beta\cos\theta)\Delta t' \boldsymbol{S}_{\text{rad}}(\boldsymbol{r},t)\mathrm{d}s\right\}$$

$$= \frac{2}{3}\left(\frac{q^2}{4\pi\varepsilon_0 c^5}\right)\left\{\frac{\ddot{r}_p^2(t')}{(1-\beta^2)^2} + \frac{[\dot{\boldsymbol{r}}_p(t')\cdot\ddot{\boldsymbol{r}}_p(t')]^2}{c^2(1-\beta^2)^3}\right\}\dot{\boldsymbol{r}}_p(t'). \quad (20)$$

At non-relativistic velocities, the second bracketed term in the above equation becomes negligible. In the special theory of relativity, the mechanical momentum of a *neutral* particle of mass $m$ is known to be $\boldsymbol{p}(t) = \gamma(t)m\boldsymbol{v}(t)$, where $\gamma(t)$ is given by $1/\sqrt{1-\beta^2} = [1-v^2(t)/c^2]^{-½}$. The time-rate of change of this mechanical momentum, $\dot{\boldsymbol{p}}(t) = \gamma(t)m\dot{\boldsymbol{v}}(t) + \gamma^3(t)m[\boldsymbol{v}(t)\cdot\dot{\boldsymbol{v}}(t)]\,\boldsymbol{v}(t)/c^2$, must then be equal to the force exerted by the external agency responsible for the acceleration (or deceleration) of the particle. In the case of charged particles, Eq.(20) indicates that an additional force will be necessary to account for the continuous shedding of EM momentum via radiation.

**5. Radiation reaction**. In the absence of acceleration (i.e., when $\boldsymbol{\alpha}=0$), the integral of the $E$-field of Eq.(16) over a small sphere of radius $R$ vanishes. This is consistent with the fact that a uniformly moving charged-particle does not radiate and, therefore, cannot be subject to radiation resistance. In general, when the velocity $\dot{\boldsymbol{r}}_p(t')$ and acceleration $\ddot{\boldsymbol{r}}_p(t')$ at the retarded time $t'$ have arbitrary magnitudes and orientations, the average of the $E$-field at time $t$ over a spherical surface of radius $R$ centered at the retarded position of the particle is given by (see Appendix E):

$\ln(1+x) = x - ½x^2 + ⅓x^3 - ¼x^4 + \cdots$

$$\boldsymbol{E}_{\text{ave}}(t) = \frac{q}{4\pi\varepsilon_0 R}\left\{\left[\frac{1-2\beta^2}{2\beta^2(1-\beta^2)} - \frac{1}{4\beta^3}\ln\left(\frac{1+\beta}{1-\beta}\right)\right]\boldsymbol{\alpha} - \left[\frac{3-2\beta^2}{2\beta^4(1-\beta^2)} - \frac{3}{4\beta^5}\ln\left(\frac{1+\beta}{1-\beta}\right)\right](\boldsymbol{\alpha}\cdot\boldsymbol{\beta})\boldsymbol{\beta}\right\}$$

$\ln[(1+x)/(1-x)] = 2(x + ⅓x^3 + ⅕x^5 + \cdots)$

$$= -\frac{q}{8\pi\varepsilon_0 R}\left[\left(\frac{1}{1-\beta^2} + \frac{1}{3} + \frac{1}{5}\beta^2 + \frac{1}{7}\beta^4 + \cdots\right)\boldsymbol{\alpha} + \left(\frac{1}{1-\beta^2} - \frac{3}{5} - \frac{3}{7}\beta^2 - \frac{3}{9}\beta^4 + \cdots\right)(\boldsymbol{\alpha}\cdot\boldsymbol{\beta})\boldsymbol{\beta}\right]$$

$$= -\frac{q}{4\pi\varepsilon_0 R}\left[\left(\frac{2}{3} + \frac{3}{5}\beta^2 + \frac{4}{7}\beta^4 + \cdots\right)\boldsymbol{\alpha} + \left(\frac{1}{5} + \frac{2}{7}\beta^2 + \frac{3}{9}\beta^4 + \cdots\right)(\boldsymbol{\alpha}\cdot\boldsymbol{\beta})\boldsymbol{\beta}\right]. \quad (21)$$

Thus, at non-relativistic velocities where $\beta$ is negligible, the spatially-averaged $E$-field over a spherical surface of radius $R$ centered at the retarded position of the accelerated point-particle will be $\boldsymbol{E}_{\text{ave}}(t) \cong -⅔q\ddot{\boldsymbol{r}}_p(t')/(4\pi\varepsilon_0 c^2 R)$. It is seen that the average $E$-field surrounding the point-charge goes to infinity as the radius $R$ of the sphere approaches zero. This, however, is not the full story, as the heretofore neglected contribution of the advanced $\delta$-function must now be taken into consideration. Returning to Eq.(13), we observe that the advanced $\delta$-function's argument vanishes when $t' = t + |\boldsymbol{r}-\boldsymbol{r}_p(t')|/c$, at which point,

$$\partial_{t'}\left\{t - t' + \sqrt{[\boldsymbol{r}-\boldsymbol{r}_p(t')]\cdot[\boldsymbol{r}-\boldsymbol{r}_p(t')]}/c\right\} = -1 - \dot{\boldsymbol{r}}_p(t')\cdot[\boldsymbol{r}-\boldsymbol{r}_p(t')]/c|\boldsymbol{r}-\boldsymbol{r}_p(t')|. \quad (22)$$

We thus have

$$\psi(\boldsymbol{r},t)=\frac{q}{4\pi\varepsilon_0\{|\boldsymbol{r}-\boldsymbol{r}_p(t')|+[\boldsymbol{r}-\boldsymbol{r}_p(t')]\cdot\dot{\boldsymbol{r}}_p(t')/c\}}. \quad (23)$$

$$\boldsymbol{A}(\boldsymbol{r},t)=\frac{\mu_0 q\,\dot{\boldsymbol{r}}_p(t')}{4\pi\{|\boldsymbol{r}-\boldsymbol{r}_p(t')|+[\boldsymbol{r}-\boldsymbol{r}_p(t')]\cdot\dot{\boldsymbol{r}}_p(t')/c\}}. \quad (24)$$

As shown in Appendix F, the advanced $\delta$-function gives rise to a second contribution to the $E$-field in the vicinity of the point-charge, which is given by

$$\boldsymbol{E}(\boldsymbol{r},t)=\frac{q}{4\pi\varepsilon_0}\left[\frac{(1-\beta^2+\boldsymbol{R}\cdot\boldsymbol{\alpha})(\hat{\boldsymbol{R}}+\boldsymbol{\beta})}{R^2(1+\beta\cos\theta)^3}-\frac{\boldsymbol{\alpha}}{R(1+\beta\cos\theta)^2}\right]. \quad (25)$$

Although the sign of $\beta$ in Eq.(25) differs from that in Eq.(16), the averaged $E$-field over a spherical surface of radius $R$ centered at the (advanced) position of the particle turns out to be the same as that in Eq.(21)—albeit with the time at which $\boldsymbol{\alpha}$ and $\boldsymbol{\beta}$ are evaluated being the *advanced* time $t'=t+|\boldsymbol{r}-\boldsymbol{r}_p(t')|/c$.

**6. The self-field**. In the limit when the radius $R$ of the sphere surrounding the particle approaches zero, the retarded time $t'_{\text{ret}}$ and the advanced time $t'_{\text{adv}}$ approach each other from the opposite sides of the present time $t$, i.e., the time at which the $E$-field is being evaluated, such that $t'_{\text{adv}}-t'_{\text{ret}}=2R/c$. In the limit when $R\to 0$, we must then take one-half of the difference between these two (space-averaged) $E$-fields in accordance with Eq.(13), in which the advanced and retarded $\delta$-functions appear with negative and positive signs, respectively. (The ½ factor accounts for the fact that only one half of each $\delta$-function contributes to the integrals being taken over $t'$ from $-\infty$ to $t$.) The actual self-field of the point-particle is thus given by

$$\boldsymbol{E}_{\text{self}}(t)=\tfrac{1}{2}\boldsymbol{E}_{\text{ave}}(t'_{\text{ret}})-\tfrac{1}{2}\boldsymbol{E}_{\text{ave}}(t'_{\text{adv}})=-(R/c)\frac{\mathrm{d}\boldsymbol{E}_{\text{ave}}}{\mathrm{d}t'}. \quad (26)$$

In Appendix G we calculate the derivative with respect to $t'$ of $\boldsymbol{E}_{\text{ave}}(t)$ given by Eq.(21). Substitution into Eq.(26) finally yields

$$\begin{aligned}\boldsymbol{E}_{\text{self}}(t)=\frac{q}{8\pi\varepsilon_0 c^3}\Big\{&\left(\frac{1}{1-\beta^2}+\frac{1}{3}+\frac{1}{5}\beta^2+\frac{1}{7}\beta^4+\frac{1}{9}\beta^6+\cdots\right)\dddot{\boldsymbol{r}}_p(t)\\&+\frac{1}{c^2}\left[\frac{1}{1-\beta^2}-3\left(\frac{1}{5}+\frac{1}{7}\beta^2+\frac{1}{9}\beta^4+\cdots\right)\right]\left[\dot{\boldsymbol{r}}_p(t)\cdot\dddot{\boldsymbol{r}}_p(t)+\ddot{\boldsymbol{r}}_p(t)\cdot\ddot{\boldsymbol{r}}_p(t)\right]\dot{\boldsymbol{r}}_p(t)\\&+\frac{1}{c^2}\left[\frac{4-2\beta^2}{(1-\beta^2)^2}-6\left(\frac{1}{5}+\frac{1}{7}\beta^2+\frac{1}{9}\beta^4+\cdots\right)\right]\left[\dot{\boldsymbol{r}}_p(t)\cdot\ddot{\boldsymbol{r}}_p(t)\right]\ddot{\boldsymbol{r}}_p(t)\\&-\frac{1}{c^4}\left[\frac{1-3\beta^2}{(1-\beta^2)^2}-15\left(\frac{1}{7}+\frac{1}{9}\beta^2+\frac{1}{11}\beta^4+\cdots\right)\right]\left[\dot{\boldsymbol{r}}_p(t)\cdot\ddot{\boldsymbol{r}}_p(t)\right]^2\dot{\boldsymbol{r}}_p(t)\Big\}.\end{aligned} \quad (27)$$

In the above equation, $\boldsymbol{\beta}=\dot{\boldsymbol{r}}_p(t)/c$ is the normalized velocity of the point-particle. Equation (27) is the *exact* expression of the self-field $\boldsymbol{E}$ acting on a point-charge $q$ moving in free space with instantaneous velocity $\dot{\boldsymbol{r}}_p(t)$, acceleration $\ddot{\boldsymbol{r}}_p(t)$, and time-rate-of-change of acceleration $\dddot{\boldsymbol{r}}_p(t)$. At non-relativistic velocities where $\beta$ is negligible, the self-field is given by $\boldsymbol{E}_{\text{self}}(t)\cong(4\pi\varepsilon_0)^{-1}(2q/3c^3)\dddot{\boldsymbol{r}}_p(t)$.

**7. Concluding remarks**. The self-field exchanges energy $\mathcal{E}$ with the point-charge $q$ at the rate of $\dot{\mathcal{E}}(t)=q\dot{\boldsymbol{r}}_p(t)\cdot\boldsymbol{E}_{\text{self}}(t)$. At non-relativistic velocities ($\beta\ll 1$), we may write

$$\begin{aligned}\Delta\mathcal{E}&=\int_{t_1}^{t_2}q\dot{\boldsymbol{r}}_p(t)\cdot\boldsymbol{E}_{\text{self}}(t)\mathrm{d}t\cong\tfrac{2}{3}\left(\frac{q^2}{4\pi\varepsilon_0 c^3}\right)\int_{t_1}^{t_2}\dot{\boldsymbol{r}}_p(t)\cdot\dddot{\boldsymbol{r}}_p(t)\mathrm{d}t\\&=\tfrac{2}{3}\left(\frac{q^2}{4\pi\varepsilon_0 c^3}\right)\left\{\left[\dot{\boldsymbol{r}}_p(t_2)\cdot\ddot{\boldsymbol{r}}_p(t_2)-\dot{\boldsymbol{r}}_p(t_1)\cdot\ddot{\boldsymbol{r}}_p(t_1)\right]-\int_{t_1}^{t_2}\ddot{r}_p^2(t)\mathrm{d}t\right\}.\end{aligned} \quad (28)$$

A comparison with the non-relativistic limit of Eq.(19) reveals that the last term in Eq.(28) accounts for the radiated energy during the $(t_1,t_2)$ interval. The remaining term, i.e., that within the square brackets, must then be the energy exchanged between the electromagnetic and non-electromagnetic components of the mass-energy of the particle. If it so happens that the product of velocity and acceleration, namely, $\dot{\boldsymbol{r}}_p\cdot\ddot{\boldsymbol{r}}_p$, at $t_2$ has returned to its initial value at $t_1$, then the action of $\boldsymbol{E}_{\text{self}}$ on the particle during the $(t_1, t_2)$ interval can be said to have resulted solely in the release of the radiated energy. This would be the case, for instance, for a particle in periodic motion with an oscillation period $T = t_2 - t_1$. Alternatively, a particle moving around a circle at constant velocity will have orthogonal velocity and acceleration vectors, resulting in $\dot{\boldsymbol{r}}_p \cdot \ddot{\boldsymbol{r}}_p = 0$. Under such circumstances, the self-field is directly related to the radiated EM energy. In general, however, any temporal variations of the product $\dot{\boldsymbol{r}}_p \cdot \ddot{\boldsymbol{r}}_p$ would indicate a mass-energy exchange between the two components (i.e., EM and non-EM) that constitute the inertial mass of the particle. The relativistic mass $mc^2/\sqrt{1-\beta^2}$, of course, varies with time as the particle undergoes acceleration and/or deceleration. It appears that the division between EM and non-EM components of the mass-energy of the particle is also a function of time, thus requiring a continuous exchange of energy between the point-particle and its associated EM field — an exchange that is presumably mediated by the action of the self-field on the particle.

The validity of the self-field as given by Eq.(27) comes under scrutiny if one considers the case of a point-charge $q$ moving around a circle of fixed radius at a constant angular velocity $\Omega$. The particle's linear velocity $\dot{\boldsymbol{r}}_p(t)$ will then have a constant magnitude $c\beta$, its acceleration $\ddot{\boldsymbol{r}}_p(t)$, being orthogonal to its velocity, will have a constant magnitude $c\beta\Omega$, and the time-derivative $\dddot{\boldsymbol{r}}_p(t)$ of its acceleration, while aligned with $-\dot{\boldsymbol{r}}_p(t)$, will have a constant magnitude $c\beta\Omega^2$. The self-field of Eq.(27) may then be simplified, as follows:

$$\boldsymbol{E}_{\text{self}}(t) = -\frac{q}{8\pi\varepsilon_0 c^3}\left(\frac{1}{1-\beta^2} + \frac{1}{3} + \frac{1}{5}\beta^2 + \frac{1}{7}\beta^4 + \frac{1}{9}\beta^6 + \cdots\right)\Omega^2\dot{\boldsymbol{r}}_p(t). \tag{29}$$

Thus, the instantaneous rate of exchange of EM energy between the field and the particle is found to be

$$\dot{\mathcal{E}}(t) = q\dot{\boldsymbol{r}}_p(t)\cdot\boldsymbol{E}_{\text{self}}(t) = -\frac{2}{3}\left(\frac{q^2}{4\pi\varepsilon_0 c^3}\right)\left(1 + \frac{9}{10}\beta^2 + \frac{12}{14}\beta^4 + \frac{15}{18}\beta^6 + \cdots\right)(c\beta\Omega)^2. \tag{30}$$

However, in accordance with Eq.(19), the instantaneous rate of radiation of EM energy must be given by

$$\frac{2}{3}\left(\frac{q^2}{4\pi\varepsilon_0 c^3}\right)\frac{(c\beta\Omega)^2}{(1-\beta^2)^2} = \frac{2}{3}\left(\frac{q^2}{4\pi\varepsilon_0 c^3}\right)[1 + 2\beta^2 + 3\beta^4 + 4\beta^6 + \cdots](c\beta\Omega)^2. \tag{31}$$

Clearly, the action of the self-field on the particle does *not* account for the radiation rate at relativistic velocities. Considering that, in the present situation, the EM and non-EM contributions to the inertial mass of the particle should be time-independent, it is difficult to fathom the discrepancy (albeit at relativistic speeds) between Eqs.(30) and (31). A plausible explanation might be that the point-particle should *not* be assumed to remain spherical at relativistic speeds; hence our Eq.(21), which has been derived by averaging the $E$-field over a spherical surface surrounding the point-charge, would lose its validity for relativistic particles.

## Appendix A

In free space, where permittivity and permeability are denoted by $\varepsilon_0$ and $\mu_0$, Maxwell's equations in the presence of free charge-density $\rho_{\text{free}}(\boldsymbol{r},t)$ and free current-density $\boldsymbol{J}_{\text{free}}(\boldsymbol{r},t)$ are written

$$\boldsymbol{\nabla}\cdot\boldsymbol{E}(\boldsymbol{r},t)=\rho_{\text{free}}(\boldsymbol{r},t)/\varepsilon_0, \tag{A1}$$

$$\boldsymbol{\nabla}\times\boldsymbol{B}(\boldsymbol{r},t)=\mu_0\boldsymbol{J}_{\text{free}}(\boldsymbol{r},t)+\mu_0\varepsilon_0\partial_t\boldsymbol{E}(\boldsymbol{r},t), \tag{A2}$$

$$\boldsymbol{\nabla}\times\boldsymbol{E}(\boldsymbol{r},t)=-\partial_t\boldsymbol{B}(\boldsymbol{r},t), \tag{A3}$$

$$\boldsymbol{\nabla}\cdot\boldsymbol{B}(\boldsymbol{r},t)=0. \tag{A4}$$

The vector potential $\boldsymbol{A}(\boldsymbol{r},t)$ is defined as the field whose curl produces the $B$-field, that is

$$\boldsymbol{\nabla}\times\boldsymbol{A}(\boldsymbol{r},t)=\boldsymbol{B}(\boldsymbol{r},t). \tag{A5}$$

Consequently, Maxwell's third equation, Eq.(A3), ensures that $\boldsymbol{E}+\partial_t\boldsymbol{A}$ is curl-free, which enables one to define the scalar potential $\psi(\boldsymbol{r},t)$ as a scalar field whose gradient (aside from an inconsequential minus sign) is equal to $\boldsymbol{E}+\partial_t\boldsymbol{A}$, that is,

$$\boldsymbol{E}(\boldsymbol{r},t)=-\boldsymbol{\nabla}\psi(\boldsymbol{r},t)-\partial_t\boldsymbol{A}(\boldsymbol{r},t). \tag{A6}$$

Substitution from Eqs.(A5) and (A6) into the remaining Maxwell equations (A1) and (A2) yields

$$\boldsymbol{\nabla}\cdot\boldsymbol{\nabla}\psi(\boldsymbol{r},t)+\partial_t\boldsymbol{\nabla}\cdot\boldsymbol{A}(\boldsymbol{r},t)=-\rho_{\text{free}}(\boldsymbol{r},t)/\varepsilon_0. \tag{A7}$$

$$\boldsymbol{\nabla}\times\boldsymbol{\nabla}\times\boldsymbol{A}(\boldsymbol{r},t)+\mu_0\varepsilon_0\partial_t^2\boldsymbol{A}(\boldsymbol{r},t)+\mu_0\varepsilon_0\partial_t\boldsymbol{\nabla}\psi(\boldsymbol{r},t)=\mu_0\boldsymbol{J}_{\text{free}}(\boldsymbol{r},t). \tag{A8}$$

Considering that $\boldsymbol{\nabla}\cdot\boldsymbol{\nabla}\psi=\boldsymbol{\nabla}^2\psi$, that $\boldsymbol{\nabla}\times\boldsymbol{\nabla}\times\boldsymbol{A}=\boldsymbol{\nabla}(\boldsymbol{\nabla}\cdot\boldsymbol{A})-\boldsymbol{\nabla}^2\boldsymbol{A}$, and that, in the Lorenz gauge, $\boldsymbol{\nabla}\cdot\boldsymbol{A}+(1/c^2)\partial_t\psi=0$, where $c=(\mu_0\varepsilon_0)^{-½}$ is the speed of light in vacuum, Eqs.(A7) and (A8) can be streamlined as follows:

$$\boldsymbol{\nabla}^2\psi(\boldsymbol{r},t)-(1/c^2)\partial_t^2\psi(\boldsymbol{r},t)=-\rho_{\text{free}}(\boldsymbol{r},t)/\varepsilon_0. \tag{A9}$$

$$\boldsymbol{\nabla}^2\boldsymbol{A}(\boldsymbol{r},t)-(1/c^2)\partial_t^2\boldsymbol{A}(\boldsymbol{r},t)=-\mu_0\boldsymbol{J}_{\text{free}}(\boldsymbol{r},t). \tag{A10}$$

Fourier transforming the sources $\rho_{\text{free}}$ and $\boldsymbol{J}_{\text{free}}$, and also the potentials $\psi$ and $\boldsymbol{A}$, we write, using the standard four-dimensional Fourier operator,

$$\rho_{\text{free}}(\boldsymbol{k},\omega)=\int_{-\infty}^{\infty}\rho_{\text{free}}(\boldsymbol{r},t)\exp[-\mathrm{i}(\boldsymbol{k}\cdot\boldsymbol{r}-\omega t)]\,\mathrm{d}\boldsymbol{r}\mathrm{d}t, \tag{A11}$$

$$\boldsymbol{J}_{\text{free}}(\boldsymbol{k},\omega)=\int_{-\infty}^{\infty}\boldsymbol{J}_{\text{free}}(\boldsymbol{r},t)\exp[-\mathrm{i}(\boldsymbol{k}\cdot\boldsymbol{r}-\omega t)]\,\mathrm{d}\boldsymbol{r}\mathrm{d}t, \tag{A12}$$

$$\psi(\boldsymbol{k},\omega)=\int_{-\infty}^{\infty}\psi(\boldsymbol{r},t)\exp[-\mathrm{i}(\boldsymbol{k}\cdot\boldsymbol{r}-\omega t)]\,\mathrm{d}\boldsymbol{r}\mathrm{d}t, \tag{A13}$$

$$\boldsymbol{A}(\boldsymbol{k},\omega)=\int_{-\infty}^{\infty}\boldsymbol{A}(\boldsymbol{r},t)\exp[-\mathrm{i}(\boldsymbol{k}\cdot\boldsymbol{r}-\omega t)]\,\mathrm{d}\boldsymbol{r}\mathrm{d}t. \tag{A14}$$

Substitution into Eqs.(A9) and (A10) now yields

$$[k^2-(\omega/c)^2]\psi(\boldsymbol{k},\omega)=\varepsilon_0^{-1}\rho_{\text{free}}(\boldsymbol{k},\omega). \tag{A15}$$

$$[k^2-(\omega/c)^2]\boldsymbol{A}(\boldsymbol{k},\omega)=\mu_0\boldsymbol{J}_{\text{free}}(\boldsymbol{k},\omega). \tag{A16}$$

At this point, one might be tempted to divide both sides of Eqs.(A15) and (A16) by $k^2-(\omega/c)^2$ in order to arrive at expressions for $\psi(\boldsymbol{k},\omega)$ and $\boldsymbol{A}(\boldsymbol{k},\omega)$. However, given the singularity at $\omega=\pm ck$, it is advisable to insert a pair of $\delta$-functions at these singular points and express the complete functional form of the potentials in the following way:

$$\psi(\boldsymbol{k},\omega)=\varepsilon_0^{-1}\rho_{\text{free}}(\boldsymbol{k},\omega)\left\{\frac{1}{k^2-(\omega/c)^2}+\left(\frac{\mathrm{i}\pi c}{2k}\right)[\delta(\omega-ck)-\delta(\omega+ck)]\right\}. \tag{A17}$$

$$\boldsymbol{A}(\boldsymbol{k},\omega)=\mu_0\boldsymbol{J}_{\text{free}}(\boldsymbol{k},\omega)\left\{\frac{1}{k^2-(\omega/c)^2}+\left(\frac{\mathrm{i}\pi c}{2k}\right)[\delta(\omega-ck)-\delta(\omega+ck)]\right\}. \tag{A18}$$

The necessity of introducing the pair of $\delta$-functions will be appreciated if one tries to evaluate either potential, say, $\psi(\boldsymbol{r},t)$, for the arbitrary source distribution $\rho(\boldsymbol{r},t)$ in the *absence* of the $\delta$-functions. One finds

$$\psi(\boldsymbol{r},t)=\mathcal{F}^{-1}\{\psi(\boldsymbol{k},\omega)\}=\frac{1}{(2\pi)^4}\int_{-\infty}^{\infty}\frac{\rho(k,\omega)\exp[\mathrm{i}(\boldsymbol{k}\cdot\boldsymbol{r}-\omega t)]}{\varepsilon_0[k^2-(\omega/c)^2]}\mathrm{d}\boldsymbol{k}\mathrm{d}\omega$$

$$=\frac{1}{(2\pi)^4\varepsilon_0}\int_{-\infty}^{\infty}\rho(\boldsymbol{r}',t')\exp[\mathrm{i}\boldsymbol{k}\cdot(\boldsymbol{r}-\boldsymbol{r}')]\left(\int_{-\infty}^{\infty}\frac{\exp[\mathrm{i}\omega(t'-t)]}{k^2-(\omega/c)^2}\mathrm{d}\omega\right)\mathrm{d}\boldsymbol{r}'\mathrm{d}t'\mathrm{d}\boldsymbol{k}. \tag{A19}$$

Evaluating the inner integral in Eq.(A19) using complex-plane integration yields

$$\int_{-\infty}^{\infty}\frac{\exp[\mathrm{i}\omega(t'-t)]}{k^2-(\omega/c)^2}\mathrm{d}\omega=-c^2\int_{-\infty}^{\infty}\frac{\exp[\mathrm{i}\omega(t'-t)]}{(\omega-ck)(\omega+ck)}\mathrm{d}\omega=\pm\mathrm{i}\pi c^2\left\{\frac{\exp[\mathrm{i}ck(t'-t)]-\exp[-\mathrm{i}ck(t'-t)]}{2ck}\right\}$$
$$=\pm(\pi c/k)\sin[ck(t-t')],\quad t\gtrless t'. \tag{A20}$$

The potential, however, should *not* depend on future times $t'>t$, so we need to augment the transfer function $1/[k^2-(\omega/c)^2]$ with the $\delta$-function pair $½\mathrm{i}(\pi c/k)[\delta(\omega-ck)-\delta(\omega+ck)]$ in order to eliminate this undesirable feature of the response function. The result is a doubling of the function appearing on the right-hand-side of Eq.(A20) for $t>t'$, and its vanishing for $t<t'$. We thus have

$$\psi(\boldsymbol{r},t)=\frac{1}{(2\pi)^4\varepsilon_0}\int_{-\infty}^{\infty}\int_{t'=-\infty}^{t}\rho(\boldsymbol{r}',t')\exp[\mathrm{i}\boldsymbol{k}\cdot(\boldsymbol{r}-\boldsymbol{r}')]\,(2\pi c/k)\sin[ck(t-t')]\,\mathrm{d}\boldsymbol{k}\mathrm{d}t'\,\mathrm{d}\boldsymbol{r}'$$
$$=\frac{c}{(2\pi)^3\varepsilon_0}\int_{-\infty}^{\infty}\int_{t'=-\infty}^{t}\rho(\boldsymbol{r}',t')k^{-1}\sin[ck(t-t')]$$
$$\times\left[\int_{\theta=0}^{\pi}2\pi k^2\sin\theta\exp(\mathrm{i}k|\boldsymbol{r}-\boldsymbol{r}'|\cos\theta)\,\mathrm{d}\theta\right]\mathrm{d}k\mathrm{d}t'\mathrm{d}\boldsymbol{r}'$$
$$=\frac{c}{(2\pi)^2\varepsilon_0}\int_{-\infty}^{\infty}\int_{t'=-\infty}^{t}[\rho(\boldsymbol{r}',t')/|\boldsymbol{r}-\boldsymbol{r}'|]\left\{\int_{k=0}^{\infty}2\sin[ck(t-t')]\sin(k|\boldsymbol{r}-\boldsymbol{r}'|)\,\mathrm{d}k\right\}\mathrm{d}t'\,\mathrm{d}\boldsymbol{r}'$$
$$=\frac{c}{(2\pi)^2\varepsilon_0}\int_{-\infty}^{\infty}\int_{t'=-\infty}^{t}[\rho(\boldsymbol{r}',t')/|\boldsymbol{r}-\boldsymbol{r}'|]$$
$$\times\int_{k=0}^{\infty}\{\cos[ck(t-t')-k|\boldsymbol{r}-\boldsymbol{r}'|]-\cos[ck(t-t')+k|\boldsymbol{r}-\boldsymbol{r}'|]\}\mathrm{d}k\mathrm{d}t'\mathrm{d}\boldsymbol{r}'$$
$$=\frac{1}{4\pi\varepsilon_0}\int_{-\infty}^{\infty}\int_{t'=-\infty}^{t}[\rho(\boldsymbol{r}',t')/|\boldsymbol{r}-\boldsymbol{r}'|][\delta(t-t'-|\boldsymbol{r}-\boldsymbol{r}'|/c)-\underbrace{\delta(t-t'+|\boldsymbol{r}-\boldsymbol{r}'|/c)}]\mathrm{d}t'\mathrm{d}\boldsymbol{r}'$$

This $\delta$-function does *not* contribute, because its argument vanishes only when $t'>t$.

$$=\frac{1}{4\pi\varepsilon_0}\int_{-\infty}^{\infty}\frac{\rho(\boldsymbol{r}',t-|\boldsymbol{r}-\boldsymbol{r}'|/c)}{|\boldsymbol{r}-\boldsymbol{r}'|}\mathrm{d}\boldsymbol{r}'. \tag{A21}$$

The final result in Eq.(A21) is the standard expression for the scalar potential in the Lorentz gauge. In the penultimate line of Eq.(A21), the "retarded" $\delta$-function has been retained while the "advanced" $\delta$-function is discarded. Nevertheless, it turns out that both $\delta$-functions contribute equally to the "self-field" of an accelerated point-charge. In the analysis of Sec.5, we have argued for the retention of the latter $\delta$-function.

## Appendix B

In order to determine the electric field $\boldsymbol{E}(\boldsymbol{r},t)$ of the point charge, we evaluate the partial derivatives of the potentials $\psi(\boldsymbol{r},t)$ and $\boldsymbol{A}(\boldsymbol{r},t)$ with respect to the spacetime coordinates. Considering the functional dependence of the potentials on the retarded time $t'$, our first step is to compute the partial derivatives of $t'$ with respect to $x$, $y$, $z$, and $t$, as follows:

$$t' = t - |\boldsymbol{r}-\boldsymbol{r}_p(t')|/c = t - \sqrt{[\boldsymbol{r}-\boldsymbol{r}_p(t')]\cdot[\boldsymbol{r}-\boldsymbol{r}_p(t')]}/c. \tag{B1}$$

$$\partial_t t' = 1 + \frac{\dot{\boldsymbol{r}}_p(t')\partial_t t' \cdot [\boldsymbol{r}-\boldsymbol{r}_p(t')]}{c|\boldsymbol{r}-\boldsymbol{r}_p(t')|} \quad \rightarrow \quad \partial_t t' = \frac{1}{1-\dot{\boldsymbol{r}}_p(t')\cdot[\boldsymbol{r}-\boldsymbol{r}_p(t')]/c|\boldsymbol{r}-\boldsymbol{r}_p(t')|}. \tag{B2}$$

$$\partial_x t' = -\frac{[\hat{\boldsymbol{x}}-\dot{\boldsymbol{r}}_p(t')\partial_x t']\cdot[\boldsymbol{r}-\boldsymbol{r}_p(t')]}{c|\boldsymbol{r}-\boldsymbol{r}_p(t')|} \quad \rightarrow \quad \partial_x t' = -\frac{x-x_p(t')}{c|\boldsymbol{r}-\boldsymbol{r}_p(t')|-\dot{\boldsymbol{r}}_p(t')\cdot[\boldsymbol{r}-\boldsymbol{r}_p(t')]}. \tag{B3}$$

$$\partial_y t' = -\frac{[\hat{\boldsymbol{y}}-\dot{\boldsymbol{r}}_p(t')\partial_y t']\cdot[\boldsymbol{r}-\boldsymbol{r}_p(t')]}{c|\boldsymbol{r}-\boldsymbol{r}_p(t')|} \quad \rightarrow \quad \partial_y t' = -\frac{y-y_p(t')}{c|\boldsymbol{r}-\boldsymbol{r}_p(t')|-\dot{\boldsymbol{r}}_p(t')\cdot[\boldsymbol{r}-\boldsymbol{r}_p(t')]}. \tag{B4}$$

$$\partial_z t' = -\frac{[\hat{\boldsymbol{z}}-\dot{\boldsymbol{r}}_p(t')\partial_z t']\cdot[\boldsymbol{r}-\boldsymbol{r}_p(t')]}{c|\boldsymbol{r}-\boldsymbol{r}_p(t')|} \quad \rightarrow \quad \partial_z t' = -\frac{z-z_p(t')}{c|\boldsymbol{r}-\boldsymbol{r}_p(t')|-\dot{\boldsymbol{r}}_p(t')\cdot[\boldsymbol{r}-\boldsymbol{r}_p(t')]}. \tag{B5}$$

Equations (14) and (15) thus yield

$$\partial_x\psi(\boldsymbol{r},t) = \frac{q}{4\pi\varepsilon_0}\frac{c\partial_x t' + [\hat{\boldsymbol{x}}-\dot{\boldsymbol{r}}_p(t')\partial_x t']\cdot\dot{\boldsymbol{r}}_p(t')/c + [\boldsymbol{r}-\boldsymbol{r}_p(t')]\cdot\ddot{\boldsymbol{r}}_p(t')\partial_x t'/c}{\{|\boldsymbol{r}-\boldsymbol{r}_p(t')| - [\boldsymbol{r}-\boldsymbol{r}_p(t')]\cdot\dot{\boldsymbol{r}}_p(t')/c\}^2}$$

$$= \frac{q}{4\pi\varepsilon_0}\frac{\{1-\dot{\boldsymbol{r}}_p(t')\cdot[\boldsymbol{r}-\boldsymbol{r}_p(t')]/c|\boldsymbol{r}-\boldsymbol{r}_p(t')|\}\dot{x}_p(t')/c - \{1-\dot{\boldsymbol{r}}_p(t')\cdot\dot{\boldsymbol{r}}_p(t')/c^2+[\boldsymbol{r}-\boldsymbol{r}_p(t')]\cdot\ddot{\boldsymbol{r}}_p(t')/c^2\}[x-x_p(t')]/|\boldsymbol{r}-\boldsymbol{r}_p(t')|}{|\boldsymbol{r}-\boldsymbol{r}_p(t')|^2\{1-[\boldsymbol{r}-\boldsymbol{r}_p(t')]\cdot\dot{\boldsymbol{r}}_p(t')/c|\boldsymbol{r}-\boldsymbol{r}_p(t')|\}^3}. \tag{B6}$$

$$\partial_t\psi(\boldsymbol{r},t) = -\frac{q}{4\pi\varepsilon_0}\times\frac{-[\boldsymbol{r}-\boldsymbol{r}_p(t')]\cdot\dot{\boldsymbol{r}}_p(t')\partial_t t'/|\boldsymbol{r}-\boldsymbol{r}_p(t')| + \dot{\boldsymbol{r}}_p(t')\cdot\dot{\boldsymbol{r}}_p(t')\partial_t t'/c - [\boldsymbol{r}-\boldsymbol{r}_p(t')]\cdot\ddot{\boldsymbol{r}}_p(t')\partial_t t'/c}{\{|\boldsymbol{r}-\boldsymbol{r}_p(t')| - [\boldsymbol{r}-\boldsymbol{r}_p(t')]\cdot\dot{\boldsymbol{r}}_p(t')/c\}^2}$$

$$= \frac{q}{4\pi\varepsilon_0}\times\frac{[\boldsymbol{r}-\boldsymbol{r}_p(t')]\cdot\dot{\boldsymbol{r}}_p(t')/|\boldsymbol{r}-\boldsymbol{r}_p(t')| - \dot{\boldsymbol{r}}_p(t')\cdot\dot{\boldsymbol{r}}_p(t')/c + [\boldsymbol{r}-\boldsymbol{r}_p(t')]\cdot\ddot{\boldsymbol{r}}_p(t')/c}{|\boldsymbol{r}-\boldsymbol{r}_p(t')|^2\{1-[\boldsymbol{r}-\boldsymbol{r}_p(t')]\cdot\dot{\boldsymbol{r}}_p(t')/c|\boldsymbol{r}-\boldsymbol{r}_p(t')|\}^3}. \tag{B7}$$

$$\partial_x\boldsymbol{A}(r,t) = \frac{\mu_0 q}{4\pi}\left\{\frac{\ddot{\boldsymbol{r}}_p(t')\partial_x t'}{|\boldsymbol{r}-\boldsymbol{r}_p(t')| - [\boldsymbol{r}-\boldsymbol{r}_p(t')]\cdot\dot{\boldsymbol{r}}_p(t')/c}\right.$$
$$\left. + \frac{\dot{\boldsymbol{r}}_p(t')\{c\partial_x t' + [\hat{\boldsymbol{x}}-\dot{\boldsymbol{r}}_p(t')\partial_x t']\cdot\dot{\boldsymbol{r}}_p(t')/c + [\boldsymbol{r}-\boldsymbol{r}_p(t')]\cdot\ddot{\boldsymbol{r}}_p(t')\partial_x t'/c\}}{\{|\boldsymbol{r}-\boldsymbol{r}_p(t')| - [\boldsymbol{r}-\boldsymbol{r}_p(t')]\cdot\dot{\boldsymbol{r}}_p(t')/c\}^2}\right\}$$

$$= \frac{\mu_0 q}{4\pi}\left\{-\frac{\ddot{\boldsymbol{r}}_p(t')[x-x_p(t')]}{c\{|r-r_p(t')| - [r-r_p(t')]\cdot\dot{r}_p(t')/c\}^2}\right.$$
$$\left. + \frac{\dot{x}_p(t')\ddot{\boldsymbol{r}}_p(t')\{|\boldsymbol{r}-\boldsymbol{r}_p(t')| - [\boldsymbol{r}-\boldsymbol{r}_p(t')]\cdot\dot{\boldsymbol{r}}_p(t')/c\} - c\,\dot{\boldsymbol{r}}_p(t')\{1-\dot{\boldsymbol{r}}_p(t')\cdot\dot{\boldsymbol{r}}_p(t')/c^2 + [\boldsymbol{r}-\boldsymbol{r}_p(t')]\cdot\ddot{\boldsymbol{r}}_p(t')/c^2\}[x-x_p(t')]}{c\{|\boldsymbol{r}-\boldsymbol{r}_p(t')| - [\boldsymbol{r}-\boldsymbol{r}_p(t')]\cdot\dot{\boldsymbol{r}}_p(t')/c\}^3}\right\}. \tag{B8}$$

$$\partial_t\boldsymbol{A}(\boldsymbol{r},t) = \frac{\mu_0 q}{4\pi}\left\{\frac{\ddot{\boldsymbol{r}}_p(t')\partial_t t'}{|\boldsymbol{r}-\boldsymbol{r}_p(t')| - [\boldsymbol{r}-\boldsymbol{r}_p(t')]\cdot\dot{\boldsymbol{r}}_p(t')/c}\right.$$
$$\left. - \frac{\dot{\boldsymbol{r}}_p(t')\{-[\boldsymbol{r}-\boldsymbol{r}_p(t')]\cdot\dot{\boldsymbol{r}}_p(t')\partial_t t'/|\boldsymbol{r}-\boldsymbol{r}_p(t')| + \dot{\boldsymbol{r}}_p(t')\cdot\dot{\boldsymbol{r}}_p(t')\partial_t t'/c - [\boldsymbol{r}-\boldsymbol{r}_p(t')]\cdot\ddot{\boldsymbol{r}}_p(t')\partial_t t'/c\}}{\{|\boldsymbol{r}-\boldsymbol{r}_p(t')| - [\boldsymbol{r}-\boldsymbol{r}_p(t')]\cdot\dot{\boldsymbol{r}}_p(t')/c\}^2}\right\}$$

$$= \frac{\mu_0 q}{4\pi}\left\{\frac{\ddot{\boldsymbol{r}}_p(t')\{|\boldsymbol{r}-\boldsymbol{r}_p(t')| - [\boldsymbol{r}-\boldsymbol{r}_p(t')]\cdot\dot{\boldsymbol{r}}_p(t')/c\}}{|\boldsymbol{r}-\boldsymbol{r}_p(t')|^2\{1-[\boldsymbol{r}-\boldsymbol{r}_p(t')]\cdot\dot{\boldsymbol{r}}_p(t')/c|\boldsymbol{r}-\boldsymbol{r}_p(t')|\}^3}\right.$$

··· continued on next page

$$+\frac{\dot{\boldsymbol r}_p(t')\{[\boldsymbol r-\boldsymbol r_p(t')]\cdot\dot{\boldsymbol r}_p(t')/|\boldsymbol r-\boldsymbol r_p(t')|-\dot{\boldsymbol r}_p(t')\cdot\dot{\boldsymbol r}_p(t')/c+[\boldsymbol r-\boldsymbol r_p(t')]\cdot\ddot{\boldsymbol r}_p(t')/c\}}{|\boldsymbol r-\boldsymbol r_p(t')|^2\{1-[\boldsymbol r-\boldsymbol r_p(t')]\cdot\dot{\boldsymbol r}_p(t')/c|\boldsymbol r-\boldsymbol r_p(t')|\}^3}\Bigg\}. \tag{B9}$$

The derivatives of $\psi(\boldsymbol r,t)$ and $\boldsymbol A(\boldsymbol r,t)$ with respect to the remaining coordinates $(y,z)$, being similar to those with respect to $x$ given by Eqs.(B6) and (B8), need not be written explicitly.

**Digression: Confirming the Lorenz gauge condition:**

$$\boldsymbol\nabla\cdot\boldsymbol A(\boldsymbol r,t)+(1/c^2)\partial_t\psi(\boldsymbol r,t)=\frac{\mu_0 q}{4\pi}\Bigg\{-\frac{\ddot{\boldsymbol r}_p(t')\cdot[\boldsymbol r-\boldsymbol r_p(t')]\{|\boldsymbol r-\boldsymbol r_p(t')|-[\boldsymbol r-\boldsymbol r_p(t')]\cdot\dot{\boldsymbol r}_p(t')/c\}}{c\{|\boldsymbol r-\boldsymbol r_p(t')|-[\boldsymbol r-\boldsymbol r_p(t')]\cdot\dot{\boldsymbol r}_p(t')/c\}^3}$$

$$+\frac{\dot{\boldsymbol r}_p(t')\cdot\dot{\boldsymbol r}_p(t')\{|\boldsymbol r-\boldsymbol r_p(t')|-[\boldsymbol r-\boldsymbol r_p(t')]\cdot\dot{\boldsymbol r}_p(t')/c\}-\{c-\dot{\boldsymbol r}_p(t')\cdot\dot{\boldsymbol r}_p(t')/c+[\boldsymbol r-\boldsymbol r_p(t')]\cdot\ddot{\boldsymbol r}_p(t')/c\}\dot{\boldsymbol r}_p(t')\cdot[\boldsymbol r-\boldsymbol r_p(t')]}{c\{|\boldsymbol r-\boldsymbol r_p(t')|-[\boldsymbol r-\boldsymbol r_p(t')]\cdot\dot{\boldsymbol r}_p(t')/c\}^3}$$

$$+\frac{[\boldsymbol r-\boldsymbol r_p(t')]\cdot\dot{\boldsymbol r}_p(t')/|\boldsymbol r-\boldsymbol r_p(t')|-\dot{\boldsymbol r}_p(t')\cdot\dot{\boldsymbol r}_p(t')/c+[\boldsymbol r-\boldsymbol r_p(t')]\cdot\ddot{\boldsymbol r}_p(t')/c}{|\boldsymbol r-\boldsymbol r_p(t')|^2\{1-[\boldsymbol r-\boldsymbol r_p(t')]\cdot\dot{\boldsymbol r}_p(t')/c|\boldsymbol r-\boldsymbol r_p(t')|\}^3}\Bigg\}=0. \tag{B10}$$

We are now in a position to derive the electric field of the point-particle with the aid of the scalar and vector potentials. Using Eqs.(B6) and (B9), the $E$-field of the traveling point-charge is found to be

$$\boldsymbol E(\boldsymbol r,t)=-\boldsymbol\nabla\psi-\partial_t\boldsymbol A$$

$$=-\frac{q}{4\pi\varepsilon_0}\frac{\{1-\dot{\boldsymbol r}_p(t')\cdot[\boldsymbol r-\boldsymbol r_p(t')]/c|\boldsymbol r-\boldsymbol r_p(t')|\}\dot{\boldsymbol r}_p(t')/c-\{1-\dot{\boldsymbol r}_p(t')\cdot\dot{\boldsymbol r}_p(t')/c^2+[\boldsymbol r-\boldsymbol r_p(t')]\cdot\ddot{\boldsymbol r}_p(t')/c^2\}[\boldsymbol r-\boldsymbol r_p(t')]/|\boldsymbol r-\boldsymbol r_p(t')|}{|\boldsymbol r-\boldsymbol r_p(t')|^2\{1-[\boldsymbol r-\boldsymbol r_p(t')]\cdot\dot{\boldsymbol r}_p(t')/c|\boldsymbol r-\boldsymbol r_p(t')|\}^3}$$

$$-\frac{\mu_0 qc^2}{4\pi}\Bigg\{\frac{\ddot{\boldsymbol r}_p(t')\{|\boldsymbol r-\boldsymbol r_p(t')|-[\boldsymbol r-\boldsymbol r_p(t')]\cdot\dot{\boldsymbol r}_p(t')/c\}/c^2}{|\boldsymbol r-\boldsymbol r_p(t')|^2\{1-[\boldsymbol r-\boldsymbol r_p(t')]\cdot\dot{\boldsymbol r}_p(t')/c|\boldsymbol r-\boldsymbol r_p(t')|\}^3}$$

$$+\frac{\{[\boldsymbol r-\boldsymbol r_p(t')]\cdot\dot{\boldsymbol r}_p(t')/c|\boldsymbol r-\boldsymbol r_p(t')|-\dot{\boldsymbol r}_p(t')\cdot\dot{\boldsymbol r}_p(t')/c^2+[\boldsymbol r-\boldsymbol r_p(t')]\cdot\ddot{\boldsymbol r}_p(t')/c^2\}\dot{\boldsymbol r}_p(t')/c}{|\boldsymbol r-\boldsymbol r_p(t')|^2\{1-[\boldsymbol r-\boldsymbol r_p(t')]\cdot\dot{\boldsymbol r}_p(t')/c|\boldsymbol r-\boldsymbol r_p(t')|\}^3}\Bigg\}$$

$$=\frac{q}{4\pi\varepsilon_0|\boldsymbol r-\boldsymbol r_p(t')|^2}\times\frac{1}{\{1-[\boldsymbol r-\boldsymbol r_p(t')]\cdot\dot{\boldsymbol r}_p(t')/c|\boldsymbol r-\boldsymbol r_p(t')|\}^3}\times$$

$$\{1-\dot{\boldsymbol r}_p(t')\cdot\dot{\boldsymbol r}_p(t')/c^2+[\boldsymbol r-\boldsymbol r_p(t')]\cdot\ddot{\boldsymbol r}_p(t')/c^2\}[\boldsymbol r-\boldsymbol r_p(t')]/|\boldsymbol r-\boldsymbol r_p(t')|$$

$$-\{1-[\boldsymbol r-\boldsymbol r_p(t')]\cdot\dot{\boldsymbol r}_p(t')/c|\boldsymbol r-\boldsymbol r_p(t')|\}\dot{\boldsymbol r}_p(t')/c$$

$$-\{1-[\boldsymbol r-\boldsymbol r_p(t')]\cdot\dot{\boldsymbol r}_p(t')/c\,|\boldsymbol r-\boldsymbol r_p(t')|\}\,|\boldsymbol r-\boldsymbol r_p(t')|\,\ddot{\boldsymbol r}_p(t')/c^2$$

$$-\{[\boldsymbol r-\boldsymbol r_p(t')]\cdot\dot{\boldsymbol r}_p(t')/c|\boldsymbol r-\boldsymbol r_p(t')|-\dot{\boldsymbol r}_p(t')\cdot\dot{\boldsymbol r}_p(t')/c^2+[\boldsymbol r-\boldsymbol r_p(t')]\cdot\ddot{\boldsymbol r}_p(t')/c^2\}\dot{\boldsymbol r}_p(t')/c$$

$$=\frac{q}{4\pi\varepsilon_0|\boldsymbol r-\boldsymbol r_p(t')|^2}\times\frac{1}{\{1-[\boldsymbol r-\boldsymbol r_p(t')]\cdot\dot{\boldsymbol r}_p(t')/c|\boldsymbol r-\boldsymbol r_p(t')|\}^3}\times$$

$$\{1-\dot{\boldsymbol r}_p(t')\cdot\dot{\boldsymbol r}_p(t')/c^2+[\boldsymbol r-\boldsymbol r_p(t')]\cdot\ddot{\boldsymbol r}_p(t')/c^2\}[\boldsymbol r-\boldsymbol r_p(t')]/|\boldsymbol r-\boldsymbol r_p(t')|$$

$$-\{1-\dot{\boldsymbol r}_p(t')\cdot\dot{\boldsymbol r}_p(t')/c^2+[\boldsymbol r-\boldsymbol r_p(t')]\cdot\ddot{\boldsymbol r}_p(t')/c^2\}\dot{\boldsymbol r}_p(t')/c$$

$$-\{1-[\boldsymbol r-\boldsymbol r_p(t')]\cdot\dot{\boldsymbol r}_p(t')/c\,|\boldsymbol r-\boldsymbol r_p(t')|\}\,|\boldsymbol r-\boldsymbol r_p(t')|\,\ddot{\boldsymbol r}_p(t')/c^2. \tag{B11}$$

Let $\boldsymbol R=\boldsymbol r-\boldsymbol r_p(t')$, and define $\boldsymbol\alpha=\ddot{\boldsymbol r}_p(t')/c^2$ and $\boldsymbol\beta=\dot{\boldsymbol r}_p(t')/c$. Defining the unit-vector $\hat{\boldsymbol R}$ as $\boldsymbol R/R$, and denoting the angle between $\boldsymbol R$ and $\boldsymbol\beta$ by $\theta$, the above expression for $\boldsymbol E(\boldsymbol r,t)$ is streamlined, as follows:

$$\boldsymbol E(\boldsymbol r,t)=\frac{q}{4\pi\varepsilon_0}\left[\frac{(1-\beta^2+\boldsymbol R\cdot\boldsymbol\alpha)(\hat{\boldsymbol R}-\boldsymbol\beta)}{R^2(1-\beta\cos\theta)^3}-\frac{\boldsymbol\alpha}{R(1-\beta\cos\theta)^2}\right]. \tag{B12}$$

Equation (B12) is the expression of the electric field of the accelerated point-charge at time $t$ and distance $\boldsymbol R$ from the location of the particle at the retarded time $t'$.

# Appendix C

Here we compute the $B$-field of the moving point-charge. From Eq.(B8) we have

$$\partial_x \boldsymbol{A}(r,t) = \frac{\mu_0 q}{4\pi}\left\{-\frac{\ddot{r}_p(t')[x - x_p(t')]}{c\{|r - r_p(t')| - [r - r_p(t')]\cdot\dot{r}_p(t')/c\}^2}\right.$$
$$\left.+\frac{\dot{r}_{px}(t')\ddot{\boldsymbol{r}}_p(t')\{|\boldsymbol{r} - \boldsymbol{r}_p(t')| - [\boldsymbol{r} - \boldsymbol{r}_p(t')]\cdot\dot{\boldsymbol{r}}_p(t')/c\} - c\,\dot{\boldsymbol{r}}_p(t')\{1 - \dot{\boldsymbol{r}}_p(t')\cdot\dot{\boldsymbol{r}}_p(t')/c^2 + [\boldsymbol{r} - \boldsymbol{r}_p(t')]\cdot\ddot{\boldsymbol{r}}_p(t')/c^2\}[x - x_p(t')]}{c\{|\boldsymbol{r} - \boldsymbol{r}_p(t')| - [\boldsymbol{r} - \boldsymbol{r}_p(t')]\cdot\dot{\boldsymbol{r}}_p(t')/c\}^3}\right\}. \tag{C1}$$

Similar expressions exist for $\partial_y \boldsymbol{A}$ and $\partial_z \boldsymbol{A}$. We thus find

$$B_x = \partial_y A_z - \partial_z A_y = \frac{\mu_0 q}{4\pi}\left\{\frac{\ddot{r}_{py}(t')[z - z_p(t')] - \ddot{z}_p(t')[y - y_p(t')]}{c\{|r - r_p(t')| - [r - r_p(t')]\cdot\dot{r}_p(t')/c\}^2}\right.$$
$$\left.+\frac{c\,\{1 - \dot{\boldsymbol{r}}_p(t')\cdot\dot{\boldsymbol{r}}_p(t')/c^2 + [\boldsymbol{r} - \boldsymbol{r}_p(t')]\cdot\ddot{\boldsymbol{r}}_p(t')/c^2\}\{\dot{y}_p(t')[z - z_p(t')] - \dot{z}_p(t')[y - y_p(t')]\}}{c\{|\boldsymbol{r} - \boldsymbol{r}_p(t')| - [\boldsymbol{r} - \boldsymbol{r}_p(t')]\cdot\dot{\boldsymbol{r}}_p(t')/c\}^3}\right\}. \tag{C2}$$

Consequently, the complete expression of the $B$-field at the observation point $(\boldsymbol{r}, t)$ is given by

$$\boldsymbol{B}(\boldsymbol{r},t) = \frac{\mu_0 q}{4\pi}\left\{\frac{\ddot{\boldsymbol{r}}_p(t')}{c\,\{|\boldsymbol{r} - \boldsymbol{r}_p(t')| - [\boldsymbol{r} - \boldsymbol{r}_p(t')]\cdot\dot{\boldsymbol{r}}_p(t')/c\}^2}\right.$$
$$\left.+\frac{c\,\{1 - \dot{\boldsymbol{r}}_p(t')\cdot\dot{\boldsymbol{r}}_p(t')/c^2 + [\boldsymbol{r} - \boldsymbol{r}_p(t')]\cdot\ddot{\boldsymbol{r}}_p(t')/c^2\}\dot{\boldsymbol{r}}_p(t')}{c\,\{|\boldsymbol{r} - \boldsymbol{r}_p(t')| - [\boldsymbol{r} - \boldsymbol{r}_p(t')]\cdot\dot{\boldsymbol{r}}_p(t')/c\}^3}\right\}\times[\boldsymbol{r} - \boldsymbol{r}_p(t')]. \tag{C3}$$

Using the parameters $\boldsymbol{R}, \widehat{\boldsymbol{R}}, \boldsymbol{\alpha}, \boldsymbol{\beta}$, and $\theta$ introduced in Sec.4, we may now write the above equation as follows:

$$\boldsymbol{B}(\boldsymbol{r},t) = \frac{q}{4\pi\varepsilon_0 c R^2}\left[\frac{(1-\beta^2 + \boldsymbol{R}\cdot\boldsymbol{\alpha})\boldsymbol{\beta}}{(1-\beta\cos\theta)^3} + \frac{R\boldsymbol{\alpha}}{(1-\beta\cos\theta)^2}\right]\times\widehat{\boldsymbol{R}}. \tag{C4}$$

A comparison with Eq.(16) reveals that $\boldsymbol{B}(\boldsymbol{r},t) = \widehat{\boldsymbol{R}}\times\boldsymbol{E}(\boldsymbol{r},t)/c$.

# Appendix D

We compute the radiation rate by integrating the Poynting vector over a spherical surface of large radius $R$. The terms that decline as $1/R^3$ and $1/R^4$ with the distance $R$ from the (retarded) position of the particle do not contribute to the radiation rate and will be ignored at the outset. The integral in Eq.(19) is thus written

$$\langle S\rangle = \oint_{\text{sphere surface}}(1-\beta\cos\theta)\boldsymbol{S}_{\text{rad}}(\boldsymbol{r},t)\cdot\mathrm{d}\boldsymbol{s}$$
$$= \left(\frac{cq^2}{16\pi^2\varepsilon_0}\right)\oint_{\text{sphere surface}}\left[\frac{(1-\beta\cos\theta)^2\alpha^2 + 2(1-\beta\cos\theta)(\boldsymbol{\alpha}\cdot\boldsymbol{\beta})(\boldsymbol{\alpha}\cdot\widehat{\boldsymbol{R}}) - (1-\beta^2)(\boldsymbol{\alpha}\cdot\widehat{\boldsymbol{R}})^2}{R^2(1-\beta\cos\theta)^5}\right]\widehat{\boldsymbol{R}}\cdot\mathrm{d}\boldsymbol{s}. \tag{D1}$$

Working in a spherical coordinate system centered at the location $\boldsymbol{r}_p(t')$ of the point-particle, with the $z$-axis aligned with the direction of $\boldsymbol{\beta} = \dot{\boldsymbol{r}}_p(t')/c$, and denoting by $(\theta, \varphi)$ the polar and azimuthal coordinates of the unit-vector $\widehat{\boldsymbol{R}}$, we will have $\widehat{\boldsymbol{R}}\cdot\mathrm{d}\boldsymbol{s} = R^2\sin\theta\,\mathrm{d}\theta\mathrm{d}\varphi$. The integrals of $\boldsymbol{\alpha}\cdot\widehat{\boldsymbol{R}}$ and $(\boldsymbol{\alpha}\cdot\widehat{\boldsymbol{R}})^2$ over the azimuthal angle $\varphi$ appearing in Eq.(D1) are readily found to be

$$\int_{\varphi=0}^{2\pi}(\boldsymbol{\alpha}\cdot\widehat{\boldsymbol{R}})\mathrm{d}\varphi = \int_{\varphi=0}^{2\pi}(\alpha_x\sin\theta\cos\varphi + \alpha_y\sin\theta\sin\varphi + \alpha_z\cos\theta)\mathrm{d}\varphi = 2\pi\alpha_z\cos\theta. \tag{D2}$$

$$\int_{\varphi=0}^{2\pi}(\boldsymbol{\alpha}\cdot\widehat{\boldsymbol{R}})^2\mathrm{d}\varphi = \int_{\varphi=0}^{2\pi}\big(\alpha_x^2\sin^2\theta\cos^2\varphi + \alpha_y^2\sin^2\theta\sin^2\varphi + \alpha_z^2\cos^2\theta$$
$$+2\alpha_x\alpha_y\sin^2\theta\sin\varphi\cos\varphi + 2\alpha_x\alpha_z\sin\theta\cos\theta\cos\varphi$$
$$+2\alpha_y\alpha_z\sin\theta\cos\theta\sin\varphi\big)\mathrm{d}\varphi = \pi(\alpha_x^2 + \alpha_y^2 - 2\alpha_z^2)\sin^2\theta + 2\pi\alpha_z^2. \tag{D3}$$

The radiation rate $\langle S \rangle$ given by Eq.(D1) may thus be expressed as follows:

$$\langle S \rangle = \left(\frac{cq^2}{16\pi^2\varepsilon_0}\right)\int_{\theta=0}^{\pi}\left\{\frac{2\pi\alpha^2\sin\theta}{(1-\beta\cos\theta)^3} + \frac{4\pi(\boldsymbol{\alpha}\cdot\boldsymbol{\beta})\alpha_z\cos\theta\sin\theta}{(1-\beta\cos\theta)^4} - \frac{(1-\beta^2)\sin\theta\left[\pi(\alpha^2-3\alpha_z^2)\sin^2\theta+2\pi\alpha_z^2\right]}{(1-\beta\cos\theta)^5}\right\}\mathrm{d}\theta. \quad \text{(D4)}$$

The integrals over the polar angle $\theta$ require one or more application of the method of integration by parts. For example, the simplest integral is found to be

$$\int_0^{\pi}\frac{\sin\theta}{(1-\beta\cos\theta)^3}\mathrm{d}\theta = -\frac{1}{2\beta(1-\beta\cos\theta)^2}\bigg|_0^{\pi} = \frac{1}{2\beta(1-\beta)^2} - \frac{1}{2\beta(1+\beta)^2} = \frac{2}{(1-\beta^2)^2}. \quad \text{(D5)}$$

All the other integrals are similarly evaluated, although the process could get somewhat tedious. The final results of these integrations are tabulated in Table D1. We thus find

$$\langle S \rangle = \left(\frac{cq^2}{16\pi^2\varepsilon_0}\right)\left[\frac{4\pi\alpha^2}{(1-\beta^2)^2} + \frac{32\pi(\boldsymbol{\alpha}\cdot\boldsymbol{\beta})^2}{3(1-\beta^2)^3} - \frac{4\pi(\alpha^2-3\alpha_z^2)}{3(1-\beta^2)^2} - \frac{4\pi(1+\beta^2)\alpha_z^2}{(1-\beta^2)^3}\right]$$

$$= \left(\frac{cq^2}{16\pi^2\varepsilon_0}\right)\left[\frac{8\pi\alpha^2}{3(1-\beta^2)^2} + \frac{32\pi(\boldsymbol{\alpha}\cdot\boldsymbol{\beta})^2}{3(1-\beta^2)^3} - \frac{8\pi(\boldsymbol{\alpha}\cdot\boldsymbol{\beta})^2}{(1-\beta^2)^3}\right] = \frac{2}{3}\left(\frac{q^2}{4\pi\varepsilon_0 c^3}\right)\left\{\frac{\ddot{r}_p^2(t')}{(1-\beta^2)^2} + \frac{\left[\dot{r}_p(t')\cdot\ddot{r}_p(t')\right]^2}{c^2(1-\beta^2)^3}\right\}. \quad \text{(D6)}$$

The rate of escape of EM momentum is computed by integrating the momentum-density $\boldsymbol{\wp}(\boldsymbol{r},t) = \boldsymbol{S}(\boldsymbol{r},t)/c^2$ over the surface of a large sphere of radius $R$. Ignoring the terms that decline as $1/R^3$ and $1/R^4$ with the radius $R$, then multiplying the resulting Poynting vector $\boldsymbol{S}_{\text{rad}}(\boldsymbol{r},t)$ by $\partial t/\partial t' = 1-\beta\cos\theta$, we find

$$\dot{\boldsymbol{p}}(t') = c^{-1}\int_{\text{sphere surface}}(1-\beta\cos\theta)\boldsymbol{S}_{\text{rad}}(\boldsymbol{r},t)\mathrm{d}s = c^{-1}\int_{\theta=0}^{\pi}\int_{\varphi=0}^{2\pi}R^2\sin\theta\,(1-\beta\cos\theta)\boldsymbol{S}_{\text{rad}}(\boldsymbol{r},t)\mathrm{d}\theta\mathrm{d}\varphi$$

$$= \left(\frac{q^2}{16\pi^2\varepsilon_0}\right)\int_{\theta=0}^{\pi}\int_{\varphi=0}^{2\pi}\sin\theta\left[\frac{(1-\beta\cos\theta)^2\alpha^2 + 2(1-\beta\cos\theta)(\boldsymbol{\alpha}\cdot\boldsymbol{\beta})(\boldsymbol{\alpha}\cdot\hat{\boldsymbol{R}}) - (1-\beta^2)(\boldsymbol{\alpha}\cdot\hat{\boldsymbol{R}})^2}{(1-\beta\cos\theta)^5}\right]\hat{\boldsymbol{R}}\mathrm{d}\theta\mathrm{d}\varphi. \quad \text{(D7)}$$

In the above equation, $\int_{\varphi=0}^{2\pi}\hat{\boldsymbol{R}}\mathrm{d}\varphi = \int_{\varphi=0}^{2\pi}(\sin\theta\cos\varphi\,\hat{\boldsymbol{x}} + \sin\theta\sin\varphi\,\hat{\boldsymbol{y}} + \cos\theta\,\hat{\boldsymbol{z}})\mathrm{d}\varphi = 2\pi\cos\theta\,\hat{\boldsymbol{z}}$. As for the remaining integrals over the azimuthal angle $\varphi$, namely, $\int_{\varphi=0}^{2\pi}(\boldsymbol{\alpha}\cdot\hat{\boldsymbol{R}})\hat{\boldsymbol{R}}\mathrm{d}\varphi$ and $\int_{\varphi=0}^{2\pi}(\boldsymbol{\alpha}\cdot\hat{\boldsymbol{R}})^2\hat{\boldsymbol{R}}\mathrm{d}\varphi$, we write

$$(\boldsymbol{\alpha}\cdot\hat{\boldsymbol{R}})\hat{\boldsymbol{R}} = (\alpha_x\sin\theta\cos\varphi + \alpha_y\sin\theta\sin\varphi + \alpha_z\cos\theta)(\sin\theta\cos\varphi\,\hat{\boldsymbol{x}} + \sin\theta\sin\varphi\,\hat{\boldsymbol{y}} + \cos\theta\,\hat{\boldsymbol{z}})$$
$$= (\alpha_x\sin^2\theta\cos^2\varphi + \alpha_y\sin^2\theta\sin\varphi\cos\varphi + \alpha_z\sin\theta\cos\theta\cos\varphi)\hat{\boldsymbol{x}}$$
$$+(\alpha_x\sin^2\theta\sin\varphi\cos\varphi + \alpha_y\sin^2\theta\sin^2\varphi + \alpha_z\sin\theta\cos\theta\sin\varphi)\hat{\boldsymbol{y}}$$
$$+(\alpha_x\sin\theta\cos\theta\cos\varphi + \alpha_y\sin\theta\cos\theta\sin\varphi + \alpha_z\cos^2\theta)\hat{\boldsymbol{z}}.$$

$$\rightarrow \int_{\varphi=0}^{2\pi}(\boldsymbol{\alpha}\cdot\hat{\boldsymbol{R}})\hat{\boldsymbol{R}}\mathrm{d}\varphi = \pi(\alpha_x\hat{\boldsymbol{x}} + \alpha_y\hat{\boldsymbol{y}})\sin^2\theta + 2\pi\alpha_z\hat{\boldsymbol{z}}\cos^2\theta = \pi(\boldsymbol{\alpha} - 3\alpha_z\hat{\boldsymbol{z}})\sin^2\theta + 2\pi\alpha_z\hat{\boldsymbol{z}}. \quad \text{(D8)}$$

$$(\boldsymbol{\alpha}\cdot\hat{\boldsymbol{R}})^2\hat{\boldsymbol{R}} = \big(\alpha_x^2\sin^2\theta\cos^2\varphi + \alpha_y^2\sin^2\theta\sin^2\varphi + \alpha_z^2\cos^2\theta + 2\alpha_x\alpha_y\sin^2\theta\sin\varphi\cos\varphi$$
$$+2\alpha_x\alpha_z\sin\theta\cos\theta\cos\varphi + 2\alpha_y\alpha_z\sin\theta\cos\theta\sin\varphi\big)(\sin\theta\cos\varphi\,\hat{\boldsymbol{x}} + \sin\theta\sin\varphi\,\hat{\boldsymbol{y}} + \cos\theta\,\hat{\boldsymbol{z}}).$$

$$\rightarrow \int_{\varphi=0}^{2\pi}(\boldsymbol{\alpha}\cdot\hat{\boldsymbol{R}})^2\hat{\boldsymbol{R}}\mathrm{d}\varphi = 2\pi(\alpha_x\hat{\boldsymbol{x}} + \alpha_y\hat{\boldsymbol{y}})\alpha_z\sin^2\theta\cos\theta + \pi\sin^2\theta\cos\theta\,(\alpha_x^2 + \alpha_y^2 - 2\alpha_z^2)\hat{\boldsymbol{z}} + 2\pi\alpha_z^2\cos\theta\,\hat{\boldsymbol{z}}$$
$$= \pi(2\alpha_z\boldsymbol{\alpha} + \alpha^2\hat{\boldsymbol{z}} - 5\alpha_z^2\hat{\boldsymbol{z}})\sin^2\theta\cos\theta + 2\pi\alpha_z^2\hat{\boldsymbol{z}}\cos\theta. \quad \text{(D9)}$$

Thus, the escape rate $\dot{\boldsymbol{p}}(t')$ of the EM linear momentum, given by Eq.(D7), is further simplified, as follows:

$$\dot{\boldsymbol{p}}(t') = \left(\frac{q^2}{16\pi^2\varepsilon_0}\right)\int_{\theta=0}^{\pi}\left\{\frac{2\pi(\boldsymbol{\alpha}\cdot\boldsymbol{\alpha})\sin\theta\cos\theta\hat{\boldsymbol{z}}}{(1-\beta\cos\theta)^3} + \frac{2\pi(\boldsymbol{\alpha}\cdot\boldsymbol{\beta})(\boldsymbol{\alpha} - 3\alpha_z\hat{\boldsymbol{z}})\sin^3\theta}{(1-\beta\cos\theta)^4} + \frac{4\pi(\boldsymbol{\alpha}\cdot\boldsymbol{\beta})\alpha_z\hat{\boldsymbol{z}}\sin\theta}{(1-\beta\cos\theta)^4}\right.$$
$$\left. - \frac{\pi(1-\beta^2)(2\alpha_z\boldsymbol{\alpha}+\alpha^2\hat{\boldsymbol{z}}-5\alpha_z^2\hat{\boldsymbol{z}})\sin^3\theta\cos\theta}{(1-\beta\cos\theta)^5} - \frac{2\pi(1-\beta^2)\alpha_z^2\hat{\boldsymbol{z}}\cos\theta\sin\theta}{(1-\beta\cos\theta)^5}\right\}\mathrm{d}\theta. \quad \text{(D10)}$$

The remaining integrals over the polar angle $\theta$ are listed in Table D1. The final result is

$$\dot{\boldsymbol{p}}(t') = \left(\frac{q^2}{4\pi\varepsilon_0}\right)\left[\frac{(\boldsymbol{\alpha}\cdot\boldsymbol{\alpha})\boldsymbol{\beta}}{(1-\beta^2)^2} + \frac{2(\boldsymbol{\alpha}\cdot\boldsymbol{\beta})(\boldsymbol{\alpha}-3\alpha_z\hat{\boldsymbol{z}})}{3(1-\beta^2)^2} + \frac{2(\beta^2+3)(\boldsymbol{\alpha}\cdot\boldsymbol{\beta})\alpha_z\hat{\boldsymbol{z}}}{3(1-\beta^2)^3} - \frac{2(\boldsymbol{\alpha}\cdot\boldsymbol{\beta})\boldsymbol{\alpha}+\alpha^2\boldsymbol{\beta}-5\alpha_z^2\boldsymbol{\beta}}{3(1-\beta^2)^2} - \frac{(5\beta+\beta^3)\alpha_z^2\hat{\boldsymbol{z}}}{3(1-\beta^2)^3}\right]$$

$$= \frac{2}{3}\left(\frac{q^2}{4\pi\varepsilon_0}\right)\left[\frac{\boldsymbol{\alpha}\cdot\boldsymbol{\alpha}}{(1-\beta^2)^2} + \frac{(\boldsymbol{\alpha}\cdot\boldsymbol{\beta})^2}{(1-\beta^2)^3}\right]\boldsymbol{\beta} = \frac{2}{3}\left(\frac{q^2}{4\pi\varepsilon_0 c^5}\right)\left\{\frac{\ddot{r}_p^2(t')}{(1-\beta^2)^2} + \frac{[\dot{\boldsymbol{r}}_p(t')\cdot\ddot{\boldsymbol{r}}_p(t')]^2}{c^2(1-\beta^2)^3}\right\}\dot{\boldsymbol{r}}_p(t'). \tag{D11}$$

Equation (D11) gives the escape rate of the EM momentum at time $t'$ for a particle of velocity $\dot{\boldsymbol{r}}_p(t')$ and acceleration $\ddot{\boldsymbol{r}}_p(t')$ at the retarded time $t'$. At non-relativistic velocities, the second bracketed term on the right-hand side of Eq.(D11) becomes negligible. Note that, if the acceleration happens to be perpendicular to the velocity (e.g., uniform circular motion), then $\dot{\boldsymbol{r}}_p(t')\cdot\ddot{\boldsymbol{r}}_p(t') = 0$. In contrast, when the acceleration and velocity are collinear — i.e., linear motion with $\ddot{\boldsymbol{r}}_p(t')$ either parallel or anti-parallel to $\dot{\boldsymbol{r}}_p(t')$ — the bracketed terms on the right-hand side of Eq.(D11) can be combined into $\ddot{r}_p^2(t')/(1-\beta^2)^3$.

| | | |
|---|---|---|
| $\int_0^\pi \frac{\sin\theta}{(1-\beta\cos\theta)^3}\mathrm{d}\theta = \frac{2}{(1-\beta^2)^2}$ | $\int_0^\pi \frac{\sin\theta\cos\theta}{(1-\beta\cos\theta)^3}\mathrm{d}\theta = \frac{2\beta}{(1-\beta^2)^2}$ | $\int_0^\pi \frac{\sin\theta}{(1-\beta\cos\theta)^4}\mathrm{d}\theta = \frac{2(\beta^2+3)}{3(1-\beta^2)^3}$ |
| $\int_0^\pi \frac{\sin\theta\cos\theta}{(1-\beta\cos\theta)^4}\mathrm{d}\theta = \frac{8\beta}{3(1-\beta^2)^3}$ | $\int_0^\pi \frac{\sin^3\theta}{(1-\beta\cos\theta)^4}\mathrm{d}\theta = \frac{4}{3(1-\beta^2)^2}$ | $\int_0^\pi \frac{\sin\theta}{(1-\beta\cos\theta)^5}\mathrm{d}\theta = \frac{2(1+\beta^2)}{(1-\beta^2)^4}$ |
| $\int_0^\pi \frac{\sin\theta\cos\theta}{(1-\beta\cos\theta)^5}\mathrm{d}\theta = \frac{2\beta^3+10\beta}{3(1-\beta^2)^4}$ | $\int_0^\pi \frac{\sin^3\theta}{(1-\beta\cos\theta)^5}\mathrm{d}\theta = \frac{4}{3(1-\beta^2)^3}$ | $\int_0^\pi \frac{\sin\theta\cos^2\theta}{(1-\beta\cos\theta)^5}\mathrm{d}\theta = \frac{2+10\beta^2}{3(1-\beta^2)^4}$ |
| $\int_0^\pi \frac{\sin^3\theta\cos\theta}{(1-\beta\cos\theta)^5}\mathrm{d}\theta = \frac{4\beta}{3(1-\beta^2)^3}$ | $\int_0^\pi \frac{\sin\theta}{(1-\beta\cos\theta)^6}\mathrm{d}\theta = \frac{2\beta^4+20\beta^2+10}{5(1-\beta^2)^5}$ | $\int_0^\pi \frac{\sin\theta\cos\theta}{(1-\beta\cos\theta)^6}\mathrm{d}\theta = \frac{20\beta+12\beta^3}{5(1-\beta^2)^5}$ |
| $\int_0^\pi \frac{\sin^3\theta}{(1-\beta\cos\theta)^6}\mathrm{d}\theta = \frac{4(\beta^2+5)}{15(1-\beta^2)^4}$ | $\int_0^\pi \frac{\sin^3\theta\cos\theta}{(1-\beta\cos\theta)^6}\mathrm{d}\theta = \frac{8\beta}{5(1-\beta^2)^4}$ | $\int_0^\pi \frac{\sin\theta\cos^3\theta}{(1-\beta\cos\theta)^6}\mathrm{d}\theta = \frac{20\beta^3+12\beta}{5(1-\beta^2)^5}$ |

**Table D1**. List of the various integrals used in the calculations leading to Eqs.(D6) and (D11).

## Appendix E

We compute the spatial average of the $E$-field over the surface of a sphere of radius $R$ at time $t$, with the particle's normalized velocity and acceleration at the retarded time being $\boldsymbol{\beta} = \dot{\boldsymbol{r}}_p(t')/c$ and $\boldsymbol{\alpha} = \ddot{\boldsymbol{r}}_p(t')/c^2$, respectively. In our chosen coordinate system, $\boldsymbol{\beta}$ is aligned with the $z$-axis. The $E$-field on the surface of the sphere of radius $R$ at time $t$ is given by Eq.(16), as follows:

$$\boldsymbol{E}(\boldsymbol{r},t) = \frac{q}{4\pi\varepsilon_0}\left[\frac{(1-\beta^2+\boldsymbol{\alpha}\cdot\boldsymbol{R})(\hat{\boldsymbol{R}}-\boldsymbol{\beta})}{R^2(1-\beta\cos\theta)^3} - \frac{\boldsymbol{\alpha}}{R(1-\beta\cos\theta)^2}\right]. \tag{E1}$$

In the absence of acceleration (i.e., $\boldsymbol{\alpha} = 0$), the integrated $E$-field over the spherical surface vanishes, as shown below. (This is consistent with the fact that the uniformly moving particle does not radiate and, therefore, cannot be subject to radiation reaction.)

$$\int_0^\pi \frac{\sin\theta(\cos\theta-\beta)}{(1-\beta\cos\theta)^3}\mathrm{d}\theta = -\frac{\cos\theta-\beta}{2\beta(1-\beta\cos\theta)^2}\bigg|_0^\pi - \frac{1}{2\beta}\int_0^\pi \frac{\sin\theta}{(1-\beta\cos\theta)^2}\mathrm{d}\theta \quad \leftarrow \text{Integration by parts}$$

$$= \frac{1+\beta}{2\beta(1+\beta)^2} + \frac{1-\beta}{2\beta(1-\beta)^2} + \frac{1}{2\beta^2(1-\beta\cos\theta)}\bigg|_0^\pi$$

$$= \frac{1}{\beta(1-\beta^2)} + \frac{1}{2\beta^2(1+\beta)} - \frac{1}{2\beta^2(1-\beta)} = \frac{1}{\beta(1-\beta^2)} - \frac{1}{\beta(1-\beta^2)} = 0. \tag{E2}$$

For an accelerated charged-particle, the last term in the expression of the $E$-field in Eq.(E1) has a well-behaved integral over a spherical surface of radius $R$, namely,

$$\int_0^\pi \frac{2\pi\boldsymbol{\alpha}\sin\theta}{(1-\beta\cos\theta)^2}\mathrm{d}\theta = -\frac{2\pi\boldsymbol{\alpha}}{\beta(1-\beta\cos\theta)}\bigg|_0^\pi = \frac{2\pi\boldsymbol{\alpha}}{\beta}\left(\frac{1}{1-\beta}-\frac{1}{1+\beta}\right) = \frac{4\pi\boldsymbol{\alpha}}{1-\beta^2}. \tag{E3}$$

The remaining term in Eq.(E1) is rather easy to evaluate when $\boldsymbol{\alpha}$ and $\boldsymbol{\beta}$ have the same orientation, in which case they make the same angle $\theta$ with the unit-vector $\hat{\boldsymbol{R}}$. Under such circumstances, the remaining integral over the spherical surface yields

$$\begin{aligned}
2\pi\alpha\int_0^\pi \frac{\sin\theta\cos\theta(\cos\theta-\beta)}{(1-\beta\cos\theta)^3}\mathrm{d}\theta &= -\frac{\pi\alpha\cos\theta(\cos\theta-\beta)}{\beta(1-\beta\cos\theta)^2}\bigg|_0^\pi - \frac{\pi\alpha}{\beta}\int_0^\pi \frac{\sin\theta(2\cos\theta-\beta)}{(1-\beta\cos\theta)^2}\mathrm{d}\theta \\
&= \frac{\pi\alpha}{\beta}\left[\frac{1-\beta}{(1-\beta)^2}-\frac{1+\beta}{(1+\beta)^2}\right] + \frac{\pi\alpha(2\cos\theta-\beta)}{\beta^2(1-\beta\cos\theta)}\bigg|_0^\pi + \frac{2\pi\alpha}{\beta^2}\int_0^\pi \frac{\sin\theta}{1-\beta\cos\theta}\mathrm{d}\theta \\
&= \frac{2\pi\alpha}{1-\beta^2} - \frac{\pi\alpha}{\beta^2}\left(\frac{2+\beta}{1+\beta}+\frac{2-\beta}{1-\beta}\right) + \frac{2\pi\alpha}{\beta^3}\ln(1-\beta\cos\theta)\bigg|_0^\pi \\
&= \frac{2\pi\alpha}{1-\beta^2} - \frac{2\pi\alpha(2-\beta^2)}{\beta^2(1-\beta^2)} + \frac{2\pi\alpha}{\beta^3}\ln\left(\frac{1+\beta}{1-\beta}\right) = -\frac{4\pi\alpha}{\beta^2} + \frac{2\pi\alpha}{\beta^3}\ln\left(\frac{1+\beta}{1-\beta}\right) \\
&= -\frac{4\pi\alpha}{\beta^2} + \frac{4\pi\alpha}{\beta^3}\left(\beta+\tfrac{1}{3}\beta^3+\tfrac{1}{5}\beta^5+\tfrac{1}{7}\beta^7+\cdots\right) \\
&= 4\pi\alpha\left(\tfrac{1}{3}+\tfrac{1}{5}\beta^2+\tfrac{1}{7}\beta^4+\cdots\right).
\end{aligned} \tag{E4}$$

$\ln(1+x) = x - \tfrac{1}{2}x^2 + \tfrac{1}{3}x^3 - \tfrac{1}{4}x^4 + \cdots$ →

Thus, when the acceleration $\boldsymbol{\alpha}$ is either parallel or anti-parallel to the velocity $\boldsymbol{\beta}$, the spatially-averaged $E$-field over a spherical surface of radius $R$ will be

$$\boldsymbol{E}_{\text{ave}}(t) = \frac{q\boldsymbol{\alpha}}{4\pi\varepsilon_0 R}\left(\tfrac{1}{3}+\tfrac{1}{5}\beta^2+\tfrac{1}{7}\beta^4+\cdots-\frac{1}{1-\beta^2}\right) = -\frac{q\boldsymbol{\alpha}}{4\pi\varepsilon_0 R}\left(\tfrac{2}{3}+\tfrac{4}{5}\beta^2+\tfrac{6}{7}\beta^4+\cdots\right). \tag{E5}$$

In general, however, $\boldsymbol{\alpha}$ and $\boldsymbol{\beta}$ will have arbitrary orientations, in which case we will have

$$\begin{aligned}
(\boldsymbol{\alpha}\cdot\hat{\boldsymbol{R}})(\hat{\boldsymbol{R}}-\boldsymbol{\beta}) &= (\alpha_x\sin\theta\cos\varphi+\alpha_y\sin\theta\sin\varphi+\alpha_z\cos\theta)[\sin\theta\cos\varphi\,\hat{\boldsymbol{x}}+\sin\theta\sin\varphi\,\hat{\boldsymbol{y}}+(\cos\theta-\beta)\hat{\boldsymbol{z}}] \\
&= (\alpha_x\sin^2\theta\cos^2\varphi+\alpha_y\sin^2\theta\sin\varphi\cos\varphi+\alpha_z\sin\theta\cos\theta\cos\varphi)\hat{\boldsymbol{x}} \\
&\quad +(\alpha_x\sin^2\theta\sin\varphi\cos\varphi+\alpha_y\sin^2\theta\sin^2\varphi+\alpha_z\sin\theta\cos\theta\sin\varphi)\hat{\boldsymbol{y}} \\
&\quad +[\alpha_x\sin\theta\,(\cos\theta-\beta)\cos\varphi+\alpha_y\sin\theta\,(\cos\theta-\beta)\sin\varphi+\alpha_z\cos\theta\,(\cos\theta-\beta)]\hat{\boldsymbol{z}}.
\end{aligned} \tag{E6}$$

$$\int_{\varphi=0}^{2\pi}(\boldsymbol{\alpha}\cdot\hat{\boldsymbol{R}})(\hat{\boldsymbol{R}}-\boldsymbol{\beta})\mathrm{d}\varphi = \pi(\alpha_x\hat{\boldsymbol{x}}+\alpha_y\hat{\boldsymbol{y}}-2\alpha_z\hat{\boldsymbol{z}})\sin^2\theta + 2\pi\alpha_z\hat{\boldsymbol{z}}(1-\beta\cos\theta). \tag{E7}$$

$$\int_{\theta=0}^{\pi}\int_{\varphi=0}^{2\pi}\frac{(\boldsymbol{\alpha}\cdot\hat{\boldsymbol{R}})(\hat{\boldsymbol{R}}-\boldsymbol{\beta})}{(1-\beta\cos\theta)^3}\sin\theta\,\mathrm{d}\varphi\,\mathrm{d}\theta = \pi(\boldsymbol{\alpha}-3\alpha_z\hat{\boldsymbol{z}})\int_0^\pi\frac{\sin^3\theta}{(1-\beta\cos\theta)^3}\mathrm{d}\theta + 2\pi\alpha_z\hat{\boldsymbol{z}}\int_0^\pi\frac{\sin\theta}{(1-\beta\cos\theta)^2}\mathrm{d}\theta. \tag{E8}$$

$$\int_0^\pi\frac{\sin\theta}{(1-\beta\cos\theta)^2}\mathrm{d}\theta = -\frac{1}{\beta(1-\beta\cos\theta)}\bigg|_0^\pi = \frac{1}{\beta(1-\beta)}-\frac{1}{\beta(1+\beta)} = \frac{2}{1-\beta^2}. \tag{E9}$$

$$\begin{aligned}
\int_0^\pi\frac{\sin^3\theta}{(1-\beta\cos\theta)^3}\mathrm{d}\theta &= -\frac{\sin^2\theta}{2\beta(1-\beta\cos\theta)^2}\bigg|_0^\pi + \frac{1}{\beta}\int_0^\pi\frac{\sin\theta\cos\theta}{(1-\beta\cos\theta)^2}\mathrm{d}\theta \\
&= -\frac{\cos\theta}{\beta^2(1-\beta\cos\theta)}\bigg|_0^\pi - \frac{1}{\beta^2}\int_0^\pi\frac{\sin\theta}{1-\beta\cos\theta}\mathrm{d}\theta = \frac{1}{\beta^2(1+\beta)}+\frac{1}{\beta^2(1-\beta)}-\frac{1}{\beta^3}\ln(1-\beta\cos\theta)\Big|_0^\pi \\
&= \frac{2}{\beta^2(1-\beta^2)} - \frac{1}{\beta^3}\ln\left(\frac{1+\beta}{1-\beta}\right).
\end{aligned} \tag{E10}$$

← Integration by parts

Consequently,

$$\boldsymbol{E}_{\text{ave}}(t)=\frac{q}{4\pi\varepsilon_0 R}\left\{¼(\boldsymbol{\alpha}-3\alpha_z\hat{\boldsymbol{z}})\left[\frac{2}{\beta^2(1-\beta^2)}-\frac{1}{\beta^3}\ln\left(\frac{1+\beta}{1-\beta}\right)\right]+\frac{\alpha_z\hat{\boldsymbol{z}}}{1-\beta^2}-\frac{\boldsymbol{\alpha}}{1-\beta^2}\right\}\leftarrow\boxed{\alpha_z\hat{\boldsymbol{z}}=(\boldsymbol{\alpha}\cdot\boldsymbol{\beta})\boldsymbol{\beta}/\beta^2}$$

$$=\frac{q}{4\pi\varepsilon_0 R}\left\{\left[\frac{1-2\beta^2}{2\beta^2(1-\beta^2)}-\frac{1}{4\beta^3}\ln\left(\frac{1+\beta}{1-\beta}\right)\right]\boldsymbol{\alpha}-\left[\frac{3-2\beta^2}{2\beta^2(1-\beta^2)}-\frac{3}{4\beta^3}\ln\left(\frac{1+\beta}{1-\beta}\right)\right]\alpha_z\hat{\boldsymbol{z}}\right\}. \tag{E11}$$

## Appendix F

In order to compute the contribution of the advanced $\delta$-function to the potentials, we return to Eq.(13) and observe that the argument of the advanced $\delta$-function vanishes when $t'=t+|\boldsymbol{r}-\boldsymbol{r}_p(t')|/c$, at which point $\partial_{t'}\{t-t'+\sqrt{[\boldsymbol{r}-\boldsymbol{r}_p(t')]\cdot[\boldsymbol{r}-\boldsymbol{r}_p(t')]}/c\}=-1-\dot{\boldsymbol{r}}_p(t')\cdot[\boldsymbol{r}-\boldsymbol{r}_p(t')]/c|\boldsymbol{r}-\boldsymbol{r}_p(t')|$. We will have

$$\psi(\boldsymbol{r},t)=\frac{q}{4\pi\varepsilon_0\{|\boldsymbol{r}-\boldsymbol{r}_p(t')|+[\boldsymbol{r}-\boldsymbol{r}_p(t')]\cdot\dot{\boldsymbol{r}}_p(t')/c\}}. \tag{F1}$$

$$\boldsymbol{A}(\boldsymbol{r},t)=\frac{\mu_0 q\,\dot{\boldsymbol{r}}_p(t')}{4\pi\{|\boldsymbol{r}-\boldsymbol{r}_p(t')|+[\boldsymbol{r}-\boldsymbol{r}_p(t')]\cdot\dot{\boldsymbol{r}}_p(t')/c\}}. \tag{F2}$$

$$t'=t+|\boldsymbol{r}-\boldsymbol{r}_p(t')|/c=t+\sqrt{[\boldsymbol{r}-\boldsymbol{r}_p(t')]\cdot[\boldsymbol{r}-\boldsymbol{r}_p(t')]}/c. \tag{F3}$$

$$\partial_t t'=1-\frac{\dot{\boldsymbol{r}}_p(t')\partial_t t'\cdot[\boldsymbol{r}-\boldsymbol{r}_p(t')]}{c|\boldsymbol{r}-\boldsymbol{r}_p(t')|}\quad\rightarrow\quad\partial_t t'=\frac{1}{1+\dot{\boldsymbol{r}}_p(t')\cdot[\boldsymbol{r}-\boldsymbol{r}_p(t')]/c|\boldsymbol{r}-\boldsymbol{r}_p(t')|}. \tag{F4}$$

$$\partial_x t'=\frac{[\hat{\boldsymbol{x}}-\dot{\boldsymbol{r}}_p(t')\partial_x t']\cdot[\boldsymbol{r}-\boldsymbol{r}_p(t')]}{c|\boldsymbol{r}-\boldsymbol{r}_p(t')|}\quad\rightarrow\quad\partial_x t'=\frac{x-x_p(t')}{c|\boldsymbol{r}-\boldsymbol{r}_p(t')|+\dot{\boldsymbol{r}}_p(t')\cdot[\boldsymbol{r}-\boldsymbol{r}_p(t')]}. \tag{F5}$$

$$\partial_y t'=\frac{[\hat{\boldsymbol{y}}-\dot{\boldsymbol{r}}_p(t')\partial_y t']\cdot[\boldsymbol{r}-\boldsymbol{r}_p(t')]}{c|\boldsymbol{r}-\boldsymbol{r}_p(t')|}\quad\rightarrow\quad\partial_y t'=\frac{y-y_p(t')}{c|\boldsymbol{r}-\boldsymbol{r}_p(t')|+\dot{\boldsymbol{r}}_p(t')\cdot[\boldsymbol{r}-\boldsymbol{r}_p(t')]}. \tag{F6}$$

$$\partial_z t'=\frac{[\hat{\boldsymbol{z}}-\dot{\boldsymbol{r}}_p(t')\partial_z t']\cdot[\boldsymbol{r}-\boldsymbol{r}_p(t')]}{c|\boldsymbol{r}-\boldsymbol{r}_p(t')|}\quad\rightarrow\quad\partial_z t'=\frac{z-z_p(t')}{c|\boldsymbol{r}-\boldsymbol{r}_p(t')|+\dot{\boldsymbol{r}}_p(t')\cdot[\boldsymbol{r}-\boldsymbol{r}_p(t')]}. \tag{F7}$$

The following analysis of the advanced potentials now parallels that of the retarded potentials in Appendix B.

$$\partial_x\psi(\boldsymbol{r},t)=-\frac{q}{4\pi\varepsilon_0}\frac{c\partial_x t'+[\hat{\boldsymbol{x}}-\dot{\boldsymbol{r}}_p(t')\partial_x t']\cdot\dot{\boldsymbol{r}}_p(t')/c+[\boldsymbol{r}-\boldsymbol{r}_p(t')]\cdot\ddot{\boldsymbol{r}}_p(t')\partial_x t'/c}{\{|\boldsymbol{r}-\boldsymbol{r}_p(t')|+[\boldsymbol{r}-\boldsymbol{r}_p(t')]\cdot\dot{\boldsymbol{r}}_p(t')/c\}^2}$$

$$=-\frac{q}{4\pi\varepsilon_0}\frac{\{1+\dot{\boldsymbol{r}}_p(t')\cdot[\boldsymbol{r}-\boldsymbol{r}_p(t')]/c|\boldsymbol{r}-\boldsymbol{r}_p(t')|\}\dot{x}_p(t')/c+\{1-\dot{\boldsymbol{r}}_p(t')\cdot\dot{\boldsymbol{r}}_p(t')/c^2+[\boldsymbol{r}-\boldsymbol{r}_p(t')]\cdot\ddot{\boldsymbol{r}}_p(t')/c^2\}[x-x_p(t')]/|\boldsymbol{r}-\boldsymbol{r}_p(t')|}{|\boldsymbol{r}-\boldsymbol{r}_p(t')|^2\{1+[\boldsymbol{r}-\boldsymbol{r}_p(t')]\cdot\dot{\boldsymbol{r}}_p(t')/c|\boldsymbol{r}-\boldsymbol{r}_p(t')|\}^3}. \tag{F8}$$

$$\partial_t\psi(\boldsymbol{r},t)=-\frac{q}{4\pi\varepsilon_0}\times\frac{-[\boldsymbol{r}-\boldsymbol{r}_p(t')]\cdot\dot{\boldsymbol{r}}_p(t')\partial_t t'/|\boldsymbol{r}-\boldsymbol{r}_p(t')|-\dot{\boldsymbol{r}}_p(t')\cdot\dot{\boldsymbol{r}}_p(t')\partial_t t'/c+[\boldsymbol{r}-\boldsymbol{r}_p(t')]\cdot\ddot{\boldsymbol{r}}_p(t')\partial_t t'/c}{\{|\boldsymbol{r}-\boldsymbol{r}_p(t')|+[\boldsymbol{r}-\boldsymbol{r}_p(t')]\cdot\dot{\boldsymbol{r}}_p(t')/c\}^2}$$

$$=\frac{q}{4\pi\varepsilon_0}\times\frac{[\boldsymbol{r}-\boldsymbol{r}_p(t')]\cdot\dot{\boldsymbol{r}}_p(t')/|\boldsymbol{r}-\boldsymbol{r}_p(t')|+\dot{\boldsymbol{r}}_p(t')\cdot\dot{\boldsymbol{r}}_p(t')/c-[\boldsymbol{r}-\boldsymbol{r}_p(t')]\cdot\ddot{\boldsymbol{r}}_p(t')/c}{|\boldsymbol{r}-\boldsymbol{r}_p(t')|^2\{1+[\boldsymbol{r}-\boldsymbol{r}_p(t')]\cdot\dot{\boldsymbol{r}}_p(t')/c|\boldsymbol{r}-\boldsymbol{r}_p(t')|\}^3}. \tag{F9}$$

$$\partial_x\boldsymbol{A}(r,t)=\frac{\mu_0 q}{4\pi}\left\{\frac{\ddot{\boldsymbol{r}}_p(t')\partial_x t'}{|\boldsymbol{r}-\boldsymbol{r}_p(t')|+[\boldsymbol{r}-\boldsymbol{r}_p(t')]\cdot\dot{\boldsymbol{r}}_p(t')/c}\right.$$

$$\left.-\frac{\dot{\boldsymbol{r}}_p(t')\{c\partial_x t'+[\hat{\boldsymbol{x}}-\dot{\boldsymbol{r}}_p(t')\partial_x t']\cdot\dot{\boldsymbol{r}}_p(t')/c+[\boldsymbol{r}-\boldsymbol{r}_p(t')]\cdot\ddot{\boldsymbol{r}}_p(t')\partial_x t'/c\}}{\{|\boldsymbol{r}-\boldsymbol{r}_p(t')|+[\boldsymbol{r}-\boldsymbol{r}_p(t')]\cdot\dot{\boldsymbol{r}}_p(t')/c\}^2}\right\}$$

$$=\frac{\mu_0 q}{4\pi}\left\{\frac{\ddot{\boldsymbol{r}}_p(t')[x-x_p(t')]}{c\{|r-r_p(t')|+[r-r_p(t')]\cdot\dot{r}_p(t')/c\}^2}\right.$$

⋯ continued on next page

$$-\frac{\dot{x}_p(t')\dot{\mathbf r}_p(t')\{|\mathbf r-\mathbf r_p(t')|+[\mathbf r-\mathbf r_p(t')]\cdot\dot{\mathbf r}_p(t')/c\}+c\,\dot{\mathbf r}_p(t')\{1-\dot{\mathbf r}_p(t')\cdot\dot{\mathbf r}_p(t')/c^2+[\mathbf r-\mathbf r_p(t')]\cdot\ddot{\mathbf r}_p(t')/c^2\}[x-x_p(t')]}{c\{|\mathbf r-\mathbf r_p(t')|+[\mathbf r-\mathbf r_p(t')]\cdot\dot{\mathbf r}_p(t')/c\}^3}\Bigg\}. \tag{F10}$$

$$\partial_t\boldsymbol A(\boldsymbol r,t)=\frac{\mu_0 q}{4\pi}\Bigg\{\frac{\ddot{\mathbf r}_p(t')\partial_t t'}{|\mathbf r-\mathbf r_p(t')|+[\mathbf r-\mathbf r_p(t')]\cdot\dot{\mathbf r}_p(t')/c}$$

$$-\frac{\dot{\mathbf r}_p(t')\{-[\mathbf r-\mathbf r_p(t')]\cdot\dot{\mathbf r}_p(t')\partial_t t'/|\mathbf r-\mathbf r_p(t')|-\dot{\mathbf r}_p(t')\cdot\dot{\mathbf r}_p(t')\partial_t t'/c+[\mathbf r-\mathbf r_p(t')]\cdot\ddot{\mathbf r}_p(t')\partial_t t'/c\}}{\{|\mathbf r-\mathbf r_p(t')|+[\mathbf r-\mathbf r_p(t')]\cdot\dot{\mathbf r}_p(t')/c\}^2}\Bigg\}$$

$$=\frac{\mu_0 q}{4\pi}\Bigg\{\frac{\ddot{\mathbf r}_p(t')\{|\mathbf r-\mathbf r_p(t')|+[\mathbf r-\mathbf r_p(t')]\cdot\dot{\mathbf r}_p(t')/c\}}{|\mathbf r-\mathbf r_p(t')|^2\{1+[\mathbf r-\mathbf r_p(t')]\cdot\dot{\mathbf r}_p(t')/c|\mathbf r-\mathbf r_p(t')|\}^3}$$

$$+\frac{\dot{\mathbf r}_p(t')\{[\mathbf r-\mathbf r_p(t')]\cdot\dot{\mathbf r}_p(t')/|\mathbf r-\mathbf r_p(t')|+\dot{\mathbf r}_p(t')\cdot\dot{\mathbf r}_p(t')/c-[\mathbf r-\mathbf r_p(t')]\cdot\ddot{\mathbf r}_p(t')/c\}}{|\mathbf r-\mathbf r_p(t')|^2\{1+[\mathbf r-\mathbf r_p(t')]\cdot\dot{\mathbf r}_p(t')/c|\mathbf r-\mathbf r_p(t')|\}^3}\Bigg\}. \tag{F11}$$

---

**Digression: Confirming the Lorenz gauge condition**:

$$\boldsymbol\nabla\cdot\boldsymbol A(\boldsymbol r,t)+(1/c^2)\partial_t\psi(\boldsymbol r,t)=\frac{\mu_0 q}{4\pi}\Bigg\{\frac{\ddot{\mathbf r}_p(t')\cdot[\mathbf r-\mathbf r_p(t')]\{|\mathbf r-\mathbf r_p(t')|+[\mathbf r-\mathbf r_p(t')]\cdot\dot{\mathbf r}_p(t')/c\}}{c\{|\mathbf r-\mathbf r_p(t')|+[\mathbf r-\mathbf r_p(t')]\cdot\dot{\mathbf r}_p(t')/c\}^3}$$

$$-\frac{\dot{\mathbf r}_p(t')\cdot\dot{\mathbf r}_p(t')\{|\mathbf r-\mathbf r_p(t')|+[\mathbf r-\mathbf r_p(t')]\cdot\dot{\mathbf r}_p(t')/c\}+\{c-\dot{\mathbf r}_p(t')\cdot\dot{\mathbf r}_p(t')/c+[\mathbf r-\mathbf r_p(t')]\cdot\ddot{\mathbf r}_p(t')/c\}\dot{\mathbf r}_p(t')\cdot[\mathbf r-\mathbf r_p(t')]}{c\{|\mathbf r-\mathbf r_p(t')|+[\mathbf r-\mathbf r_p(t')]\cdot\dot{\mathbf r}_p(t')/c\}^3}$$

$$+\frac{[\mathbf r-\mathbf r_p(t')]\cdot\dot{\mathbf r}_p(t')/|\mathbf r-\mathbf r_p(t')|+\dot{\mathbf r}_p(t')\cdot\dot{\mathbf r}_p(t')/c-[\mathbf r-\mathbf r_p(t')]\cdot\ddot{\mathbf r}_p(t')/c}{|\mathbf r-\mathbf r_p(t')|^2\{1+[\mathbf r-\mathbf r_p(t')]\cdot\dot{\mathbf r}_p(t')/c|\mathbf r-\mathbf r_p(t')|\}^3}\Bigg\}=0. \tag{F12}$$

---

We are now in a position to compute the $E$-field produced by the advanced potentials of Eqs.(F1) and (F2), which correspond to the advanced time $t'$ given in Eq.(F3), as follows:

$$\boldsymbol E(\boldsymbol r,t)=-\boldsymbol\nabla\psi-\partial_t\boldsymbol A$$

$$=\frac{q}{4\pi\varepsilon_0}\frac{\{1+\dot{\mathbf r}_p(t')\cdot[\mathbf r-\mathbf r_p(t')]/c|\mathbf r-\mathbf r_p(t')|\}\ddot{\mathbf r}_p(t')/c+\{1-\dot{\mathbf r}_p(t')\cdot\dot{\mathbf r}_p(t')/c^2+[\mathbf r-\mathbf r_p(t')]\cdot\ddot{\mathbf r}_p(t')/c^2\}[\mathbf r-\mathbf r_p(t')]/|\mathbf r-\mathbf r_p(t')|}{|\mathbf r-\mathbf r_p(t')|^2\{1+[\mathbf r-\mathbf r_p(t')]\cdot\dot{\mathbf r}_p(t')/c|\mathbf r-\mathbf r_p(t')|\}^3}$$

$$-\frac{\mu_0 qc^2}{4\pi}\Bigg\{\frac{\ddot{\mathbf r}_p(t')\{|\mathbf r-\mathbf r_p(t')|+[\mathbf r-\mathbf r_p(t')]\cdot\dot{\mathbf r}_p(t')/c\}/c^2}{|\mathbf r-\mathbf r_p(t')|^2\{1+[\mathbf r-\mathbf r_p(t')]\cdot\dot{\mathbf r}_p(t')/c|\mathbf r-\mathbf r_p(t')|\}^3}$$

$$+\frac{\{[\mathbf r-\mathbf r_p(t')]\cdot\dot{\mathbf r}_p(t')/c|\mathbf r-\mathbf r_p(t')|+\dot{\mathbf r}_p(t')\cdot\dot{\mathbf r}_p(t')/c^2-[\mathbf r-\mathbf r_p(t')]\cdot\ddot{\mathbf r}_p(t')/c^2\}\dot{\mathbf r}_p(t')/c}{|\mathbf r-\mathbf r_p(t')|^2\{1+[\mathbf r-\mathbf r_p(t')]\cdot\dot{\mathbf r}_p(t')/c|\mathbf r-\mathbf r_p(t')|\}^3}\Bigg\}$$

$$=\frac{q}{4\pi\varepsilon_0|\mathbf r-\mathbf r_p(t')|^2}\times\frac{1}{\{1+[\mathbf r-\mathbf r_p(t')]\cdot\dot{\mathbf r}_p(t')/c|\mathbf r-\mathbf r_p(t')|\}^3}\times$$

$$\{1-\dot{\mathbf r}_p(t')\cdot\dot{\mathbf r}_p(t')/c^2+[\mathbf r-\mathbf r_p(t')]\cdot\ddot{\mathbf r}_p(t')/c^2\}[\mathbf r-\mathbf r_p(t')]/|\mathbf r-\mathbf r_p(t')|$$

$$+\{1+[\mathbf r-\mathbf r_p(t')]\cdot\dot{\mathbf r}_p(t')/c|\mathbf r-\mathbf r_p(t')|\}\ddot{\mathbf r}_p(t')/c$$

$$-\{1+[\mathbf r-\mathbf r_p(t')]\cdot\dot{\mathbf r}_p(t')/c\,|\mathbf r-\mathbf r_p(t')|\}\,|\mathbf r-\mathbf r_p(t')|\,\ddot{\mathbf r}_p(t')/c^2$$

$$-\{[\mathbf r-\mathbf r_p(t')]\cdot\dot{\mathbf r}_p(t')/c|\mathbf r-\mathbf r_p(t')|+\dot{\mathbf r}_p(t')\cdot\dot{\mathbf r}_p(t')/c^2-[\mathbf r-\mathbf r_p(t')]\cdot\ddot{\mathbf r}_p(t')/c^2\}\dot{\mathbf r}_p(t')/c$$

··· continued on next page

$$= \frac{q}{4\pi\varepsilon_0|\boldsymbol{r}-\boldsymbol{r}_p(t')|^2} \times \frac{1}{\{1+[\boldsymbol{r}-\boldsymbol{r}_p(t')]\cdot\dot{\boldsymbol{r}}_p(t')/c|\boldsymbol{r}-\boldsymbol{r}_p(t')|\}^3} \times$$

$$\{1-\dot{\boldsymbol{r}}_p(t')\cdot\dot{\boldsymbol{r}}_p(t')/c^2+[\boldsymbol{r}-\boldsymbol{r}_p(t')]\cdot\ddot{\boldsymbol{r}}_p(t')/c^2\}[\boldsymbol{r}-\boldsymbol{r}_p(t')]/|\boldsymbol{r}-\boldsymbol{r}_p(t')|$$

$$+\{1-\dot{\boldsymbol{r}}_p(t')\cdot\dot{\boldsymbol{r}}_p(t')/c^2+[\boldsymbol{r}-\boldsymbol{r}_p(t')]\cdot\ddot{\boldsymbol{r}}_p(t')/c^2\}\dot{\boldsymbol{r}}_p(t')/c$$

$$-\{1+[\boldsymbol{r}-\boldsymbol{r}_p(t')]\cdot\dot{\boldsymbol{r}}_p(t')/c\,|\boldsymbol{r}-\boldsymbol{r}_p(t')|\}\,|\boldsymbol{r}-\boldsymbol{r}_p(t')|\,\ddot{\boldsymbol{r}}_p(t')/c^2. \tag{F13}$$

As before, we let $\boldsymbol{R}=\boldsymbol{r}-\boldsymbol{r}_p(t')$, $\boldsymbol{\alpha}=\ddot{\boldsymbol{r}}_p(t')/c^2$, $\boldsymbol{\beta}=\dot{\boldsymbol{r}}_p(t')/c$, and denote by $\theta$ the angle between $\boldsymbol{R}$ and $\boldsymbol{\beta}$, so that the expression in Eq.(F13) may be rewritten as follows:

$$\boldsymbol{E}(\boldsymbol{r},t)=\frac{q}{4\pi\varepsilon_0}\left[\frac{(1-\beta^2+\boldsymbol{R}\cdot\boldsymbol{\alpha})(\hat{\boldsymbol{R}}+\boldsymbol{\beta})}{R^2(1+\beta\cos\theta)^3}-\frac{\boldsymbol{\alpha}}{R(1+\beta\cos\theta)^2}\right]. \tag{F14}$$

Although the sign of $\beta$ in Eq.(F14) is the opposite of that in Eq.(16), the spatially averaged $E$-field over a spherical surface of radius $R$ centered at the (advanced) position of the particle turns out to be the same as that in Eq.(21) — albeit with the time $t'$ at which $\boldsymbol{\alpha}$ and $\boldsymbol{\beta}$ are evaluated being the *advanced* time, which is given by Eq.(F3).

## Appendix G

We compute the derivative of the spatially-averaged $E$-field over a small spherical surface of radius $R$ surrounding the point-particle; see Eq.(E11). The derivative is taken with respect to the retarded time $t'$.

$$\boldsymbol{E}_{\text{ave}}(t)=\frac{q}{4\pi\varepsilon_0 R}\left\{\left[\boldsymbol{\alpha}-\frac{3(\boldsymbol{\alpha}\cdot\boldsymbol{\beta})\boldsymbol{\beta}}{\beta^2}\right]\left[\frac{1}{2\beta^2(1-\beta^2)}-\frac{1}{4\beta^3}\ln\left(\frac{1+\beta}{1-\beta}\right)\right]-\frac{\boldsymbol{\alpha}-(\boldsymbol{\alpha}\cdot\boldsymbol{\beta})\boldsymbol{\beta}/\beta^2}{1-\beta^2}\right\}. \tag{G1}$$

$$\frac{\mathrm{d}\beta}{\mathrm{d}t'}=\frac{\mathrm{d}}{\mathrm{d}t'}(\boldsymbol{\beta}\cdot\boldsymbol{\beta})^{1/2}=\tfrac{1}{2}(\dot{\boldsymbol{\beta}}\cdot\boldsymbol{\beta}+\boldsymbol{\beta}\cdot\dot{\boldsymbol{\beta}})(\boldsymbol{\beta}\cdot\boldsymbol{\beta})^{-1/2}=\boldsymbol{\beta}\cdot\dot{\boldsymbol{\beta}}/\beta=c\boldsymbol{\alpha}\cdot\boldsymbol{\beta}/\beta. \tag{G2}$$

$$\frac{\mathrm{d}}{\mathrm{d}t'}\left[\frac{(\boldsymbol{\alpha}\cdot\boldsymbol{\beta})\boldsymbol{\beta}}{\beta^2}\right]=\frac{(\dot{\boldsymbol{\alpha}}\cdot\boldsymbol{\beta}+c\alpha^2)\boldsymbol{\beta}+c(\boldsymbol{\alpha}\cdot\boldsymbol{\beta})\boldsymbol{\alpha}}{\beta^2}-\frac{2(\boldsymbol{\beta}\cdot\dot{\boldsymbol{\beta}})(\boldsymbol{\alpha}\cdot\boldsymbol{\beta})\boldsymbol{\beta}}{\beta^4}=\frac{(\dot{\boldsymbol{\alpha}}\cdot\boldsymbol{\beta})\boldsymbol{\beta}}{\beta^2}+\frac{c(\boldsymbol{\alpha}\cdot\boldsymbol{\beta})\boldsymbol{\alpha}}{\beta^2}+\frac{c[\alpha^2\beta^2-2(\boldsymbol{\alpha}\cdot\boldsymbol{\beta})^2]\boldsymbol{\beta}}{\beta^4}. \tag{G3}$$

$$\frac{\mathrm{d}\boldsymbol{E}_{\text{ave}}}{\mathrm{d}t'}=\left\{\dot{\boldsymbol{\alpha}}-\frac{3(\dot{\boldsymbol{\alpha}}\cdot\boldsymbol{\beta})\boldsymbol{\beta}}{\beta^2}-\frac{3c(\boldsymbol{\alpha}\cdot\boldsymbol{\beta})\boldsymbol{\alpha}}{\beta^2}-\frac{3c[\alpha^2\beta^2-2(\boldsymbol{\alpha}\cdot\boldsymbol{\beta})^2]\boldsymbol{\beta}}{\beta^4}\right\}\left[\frac{1}{2\beta^2(1-\beta^2)}-\frac{1}{4\beta^3}\ln\left(\frac{1+\beta}{1-\beta}\right)\right]$$

Temporarily dropping the coefficient $q/(4\pi\varepsilon_0 R)$.

$$+\left[\boldsymbol{\alpha}-\frac{3(\boldsymbol{\alpha}\cdot\boldsymbol{\beta})\boldsymbol{\beta}}{\beta^2}\right]\left[-\frac{\boldsymbol{\beta}\cdot\dot{\boldsymbol{\beta}}(1-2\beta^2)}{\beta^4(1-\beta^2)^2}+\frac{3\boldsymbol{\beta}\cdot\dot{\boldsymbol{\beta}}}{4\beta^5}\ln\left(\frac{1+\beta}{1-\beta}\right)-\frac{1}{4\beta^3}\left(\frac{\boldsymbol{\beta}\cdot\dot{\boldsymbol{\beta}}/\beta}{1+\beta}+\frac{\boldsymbol{\beta}\cdot\dot{\boldsymbol{\beta}}/\beta}{1-\beta}\right)\right]$$

$$-\frac{1}{1-\beta^2}\left\{\dot{\boldsymbol{\alpha}}-\frac{(\dot{\boldsymbol{\alpha}}\cdot\boldsymbol{\beta})\boldsymbol{\beta}}{\beta^2}-\frac{c(\boldsymbol{\alpha}\cdot\boldsymbol{\beta})\boldsymbol{\alpha}}{\beta^2}-\frac{c[\alpha^2\beta^2-2(\boldsymbol{\alpha}\cdot\boldsymbol{\beta})^2]\boldsymbol{\beta}}{\beta^4}\right\}-\frac{2\boldsymbol{\beta}\cdot\dot{\boldsymbol{\beta}}[\boldsymbol{\alpha}-(\boldsymbol{\alpha}\cdot\boldsymbol{\beta})\boldsymbol{\beta}/\beta^2]}{(1-\beta^2)^2}$$

$$\frac{\mathrm{d}\boldsymbol{E}_{\text{ave}}}{\mathrm{d}t'}=\left\{\dot{\boldsymbol{\alpha}}-\frac{3(\dot{\boldsymbol{\alpha}}\cdot\boldsymbol{\beta})\boldsymbol{\beta}}{\beta^2}-\frac{3c(\boldsymbol{\alpha}\cdot\boldsymbol{\beta})\boldsymbol{\alpha}}{\beta^2}-\frac{3c[\alpha^2\beta^2-2(\boldsymbol{\alpha}\cdot\boldsymbol{\beta})^2]\boldsymbol{\beta}}{\beta^4}\right\}\left[\frac{1}{2\beta^2(1-\beta^2)}-\frac{1}{4\beta^3}\ln\left(\frac{1+\beta}{1-\beta}\right)\right]$$

$$+\left[\boldsymbol{\alpha}-\frac{3(\boldsymbol{\alpha}\cdot\boldsymbol{\beta})\boldsymbol{\beta}}{\beta^2}\right]\left[\frac{5\beta^2-3}{2\beta^3(1-\beta^2)^2}+\frac{3}{4\beta^4}\ln\left(\frac{1+\beta}{1-\beta}\right)\right]\frac{c\boldsymbol{\alpha}\cdot\boldsymbol{\beta}}{\beta}$$

$$-\frac{\beta^4\dot{\boldsymbol{\alpha}}-\beta^2(\dot{\boldsymbol{\alpha}}\cdot\boldsymbol{\beta})\boldsymbol{\beta}-c\beta^2(\boldsymbol{\alpha}\cdot\boldsymbol{\beta})\boldsymbol{\alpha}-c[\alpha^2\beta^2-2(\boldsymbol{\alpha}\cdot\boldsymbol{\beta})^2]\boldsymbol{\beta}}{\beta^4(1-\beta^2)}-\frac{2c\boldsymbol{\alpha}\cdot\boldsymbol{\beta}[\boldsymbol{\alpha}-(\boldsymbol{\alpha}\cdot\boldsymbol{\beta})\boldsymbol{\beta}/\beta^2]}{(1-\beta^2)^2}$$

··· continued on next page

$$\frac{\mathrm{d}\boldsymbol{E}_{\text{ave}}}{\mathrm{d}t'} = \dot{\boldsymbol{\alpha}}\left[\frac{1-2\beta^2}{2\beta^2(1-\beta^2)} - \frac{1}{4\beta^3}\ln\left(\frac{1+\beta}{1-\beta}\right)\right] - \frac{(\dot{\boldsymbol{\alpha}}\cdot\boldsymbol{\beta})\boldsymbol{\beta}}{\beta^2}\left[\frac{3-2\beta^2}{2\beta^2(1-\beta^2)} - \frac{3}{4\beta^3}\ln\left(\frac{1+\beta}{1-\beta}\right)\right]$$

$$-\left\{\frac{3c(\boldsymbol{\alpha}\cdot\boldsymbol{\beta})\boldsymbol{\alpha}}{\beta^2} + \frac{3c[\alpha^2\beta^2-2(\boldsymbol{\alpha}\cdot\boldsymbol{\beta})^2]\boldsymbol{\beta}}{\beta^4}\right\}\left[\frac{1}{2\beta^2(1-\beta^2)} - \frac{1}{4\beta^3}\ln\left(\frac{1+\beta}{1-\beta}\right)\right] + \frac{c\beta^2(\boldsymbol{\alpha}\cdot\boldsymbol{\beta})\boldsymbol{\alpha} + c[\alpha^2\beta^2 - 2(\boldsymbol{\alpha}\cdot\boldsymbol{\beta})^2]\boldsymbol{\beta}}{\beta^4(1-\beta^2)}$$

$$+\left[\boldsymbol{\alpha} - \frac{3(\boldsymbol{\alpha}\cdot\boldsymbol{\beta})\boldsymbol{\beta}}{\beta^2}\right]\left[\frac{5\beta^2-3}{2\beta^3(1-\beta^2)^2} + \frac{3}{4\beta^4}\ln\left(\frac{1+\beta}{1-\beta}\right)\right]\frac{c\boldsymbol{\alpha}\cdot\boldsymbol{\beta}}{\beta} - \frac{2c\boldsymbol{\alpha}\cdot\boldsymbol{\beta}[\boldsymbol{\alpha} - (\boldsymbol{\alpha}\cdot\boldsymbol{\beta})\boldsymbol{\beta}/\beta^2]}{(1-\beta^2)^2}$$

$\ln[(1+x)/(1-x)] = 2(x + ⅓x^3 + ⅕x^5 + \cdots)$

$$\frac{\mathrm{d}\boldsymbol{E}_{\text{ave}}}{\mathrm{d}t'} = -½\dot{\boldsymbol{\alpha}}\left(\frac{1}{1-\beta^2} + \frac{1}{3} + \frac{1}{5}\beta^2 + \frac{1}{7}\beta^4 + \cdots\right) - \frac{(\dot{\boldsymbol{\alpha}}\cdot\boldsymbol{\beta})\boldsymbol{\beta}}{2\beta^2}\left[\frac{1}{1-\beta^2} - 3\left(\frac{1}{3} + \frac{1}{5}\beta^2 + \frac{1}{7}\beta^4 + \cdots\right)\right]$$

$$-\frac{3}{2}\left\{\frac{c(\boldsymbol{\alpha}\cdot\boldsymbol{\beta})\boldsymbol{\alpha}}{\beta^2} + \frac{c[\alpha^2\beta^2-2(\boldsymbol{\alpha}\cdot\boldsymbol{\beta})^2]\boldsymbol{\beta}}{\beta^4}\right\}\left[\frac{1}{3(1-\beta^2)} - \left(\frac{1}{3} + \frac{1}{5}\beta^2 + \frac{1}{7}\beta^4 + \cdots\right)\right]$$

$$+½\left[\boldsymbol{\alpha} - \frac{3(\boldsymbol{\alpha}\cdot\boldsymbol{\beta})\boldsymbol{\beta}}{\beta^2}\right]\left[\frac{1+\beta^2}{(1-\beta^2)^2} + 3\left(\frac{1}{5} + \frac{1}{7}\beta^2 + \frac{1}{9}\beta^4 + \cdots\right)\right]c\boldsymbol{\alpha}\cdot\boldsymbol{\beta} - \frac{2c\boldsymbol{\alpha}\cdot\boldsymbol{\beta}[\boldsymbol{\alpha} - (\boldsymbol{\alpha}\cdot\boldsymbol{\beta})\boldsymbol{\beta}/\beta^2]}{(1-\beta^2)^2}$$

$$\frac{\mathrm{d}\boldsymbol{E}_{\text{ave}}}{\mathrm{d}t'} = -½\dot{\boldsymbol{\alpha}}\left(\frac{1}{1-\beta^2} + \frac{1}{3} + \frac{1}{5}\beta^2 + \frac{1}{7}\beta^4 + \cdots\right) - ½(\dot{\boldsymbol{\alpha}}\cdot\boldsymbol{\beta})\boldsymbol{\beta}\left[\frac{1}{1-\beta^2} - 3\left(\frac{1}{5} + \frac{1}{7}\beta^2 + \frac{1}{9}\beta^4 + \cdots\right)\right]$$

$$-½\left[c(\boldsymbol{\alpha}\cdot\boldsymbol{\beta})\boldsymbol{\alpha} + c\alpha^2\boldsymbol{\beta} - \frac{2c(\boldsymbol{\alpha}\cdot\boldsymbol{\beta})^2\boldsymbol{\beta}}{\beta^2}\right]\left[\frac{1}{1-\beta^2} - 3\left(\frac{1}{5} + \frac{1}{7}\beta^2 + \frac{1}{9}\beta^4 + \cdots\right)\right]$$

$$+½\left[c(\boldsymbol{\alpha}\cdot\boldsymbol{\beta})\boldsymbol{\alpha} - \frac{3c(\boldsymbol{\alpha}\cdot\boldsymbol{\beta})^2\boldsymbol{\beta}}{\beta^2}\right]\left[\frac{1+\beta^2}{(1-\beta^2)^2} + 3\left(\frac{1}{5} + \frac{1}{7}\beta^2 + \frac{1}{9}\beta^4 + \cdots\right)\right] - \frac{2c(\boldsymbol{\alpha}\cdot\boldsymbol{\beta})\boldsymbol{\alpha}}{(1-\beta^2)^2} + \frac{2c(\boldsymbol{\alpha}\cdot\boldsymbol{\beta})^2\boldsymbol{\beta}/\beta^2}{(1-\beta^2)^2}$$

$$\frac{\mathrm{d}\boldsymbol{E}_{\text{ave}}}{\mathrm{d}t'} = -\frac{q}{8\pi\varepsilon_0 R}\left\{\left(\frac{1}{1-\beta^2} + \frac{1}{3} + \frac{1}{5}\beta^2 + \frac{1}{7}\beta^4 + \frac{1}{9}\beta^6 + \cdots\right)\dot{\boldsymbol{\alpha}}\right.$$

Restoring the coefficient $q/(4\pi\varepsilon_0 R)$.

$$+\left[\frac{1}{1-\beta^2} - 3\left(\frac{1}{5} + \frac{1}{7}\beta^2 + \frac{1}{9}\beta^4 + \cdots\right)\right](\dot{\boldsymbol{\alpha}}\cdot\boldsymbol{\beta} + c\boldsymbol{\alpha}\cdot\boldsymbol{\alpha})\boldsymbol{\beta}$$

$$+\left[\frac{4-2\beta^2}{(1-\beta^2)^2} - 6\left(\frac{1}{5} + \frac{1}{7}\beta^2 + \frac{1}{9}\beta^4 + \cdots\right)\right]c(\boldsymbol{\alpha}\cdot\boldsymbol{\beta})\boldsymbol{\alpha}$$

$$\left.-\left[\frac{1-3\beta^2}{(1-\beta^2)^2} - 15\left(\frac{1}{7} + \frac{1}{9}\beta^2 + \frac{1}{11}\beta^4 + \cdots\right)\right]c(\boldsymbol{\alpha}\cdot\boldsymbol{\beta})^2\boldsymbol{\beta}\right\}. \qquad \text{(G4)}$$

At non-relativistic velocities, where $\beta$ is negligible, we will have $\mathrm{d}\boldsymbol{E}_{\text{ave}}/\mathrm{d}t' \cong -⅔\, q\ddot{\boldsymbol{r}}_p(t')/(4\pi\varepsilon_0 c^2 R)$.